\documentclass[twocolumn,showpacs,superscriptaddress,10pt]{revtex4-1} 
\usepackage{amsmath,amssymb,graphicx}
\usepackage{psfrag}
\usepackage{epstopdf}
\usepackage{color}
\newcommand{\be} {\begin{equation}}
\newcommand{\ee} {\end{equation}}

\begin{document}
\title{Anomalous optical bistability and robust entanglement of mechanical oscillators using two-photon coherence}
\author{Eyob A. Sete}
\affiliation{Department of Electrical Engineering, University of California, Riverside, California 92521, USA}
\email{Corresponding author: esete@ee.ucr.edu}
\author{H. Eleuch}
\affiliation{Department of Physics, McGill University, Montreal, Canada H3A 2T8}

\begin{abstract}
We analyze the optical bistability and the entanglement of two movable mirrors coupled to a two-mode laser in a doubly-resonant cavity. We show that in stark contrast to the usual red-detuned condition to observe bistability in single-mode optomechanics, the optical intensities exhibit bistability for all values of cavity-laser detuning due to intermode coupling induced by the two-photon atomic coherence. Interestingly, an unconventional bistability with ``ribbon''-shaped hysteresis can be observed for certain range of cavity-laser detuning. We also demonstrate that the atomic coherence leads to a strong entanglement between the movable mirrors in the adiabatic regime. Surprisingly, the mirror-mirror entanglement is shown to persist for environment temperature of the phonon bath up to $12~\text{K}$ using experimental parameters.
\end{abstract}
\maketitle

\section{Introduction}
The entanglement of macroscopic systems provides insight into the fundamental questions regarding the quantum to classical transitions. In this respect, mechanical oscillators are of particular interest because of their resemblance to prototypical classical systems. In addition to the theoretical proposals \cite{Man02,Mar09,Abd12,Gua14,Set14} that predict entanglement between a mechanical oscillator and a cavity field, the recent experimental realization \cite{Leh13} of entanglement between the motion of a mechanical oscillator and a propagating microwave in an electromechanical circuit makes optomechanical coupling a promising platform for generating macroscopic entanglement. Other interesting theoretical proposals include the entanglement of the mirrors of two different cavities illuminated by entangled light beams \cite{Zha03}, and the entanglement of two mirrors of a double-cavity set up coupled to squeezed light \cite{Pin05,Hua09,Set14}. Optomechanical coupling is also shown to exhibit nonlinear effects such as squeezing \cite{Set12,Fab94,Woo08,Pur13}, optical bistability \cite{Set12,Tre96,Dor83,Mey85,Goz85,Jia12,Kyr13}, optomechanically induced transparency \cite{Wei10,Saf11}, and photon blockade \cite{Rab11,Nun11}, among others.

A two-mode laser with a gain medium containing an ensemble of three-level atoms in a cascade configuration is shown to exhibit quenching of spontaneous emission \cite{Scu85} and squeezed light \cite{Scu82,Set07,Ale07d} due to the two-photon coherence between the upper and lower level of the atoms. In such a laser, the two-photon coherence can be generated in two ways: either by injecting the atoms in a coherent superposition of the upper and lower levels of each atom (\textit{injected coherence}), or coupling the same levels with a strong laser (\textit{driven coherence}). These coherences are shown to generate entanglement between the cavity modes of the laser \cite{Zub05,Ale07a,Ale07c,Set08}, and more recently to entangle the movable mirrors of the doubly-resonant cavity \cite{Zho11,Zub13,Zub13-1}.
\begin{figure*}[t]
\includegraphics[width=15cm]{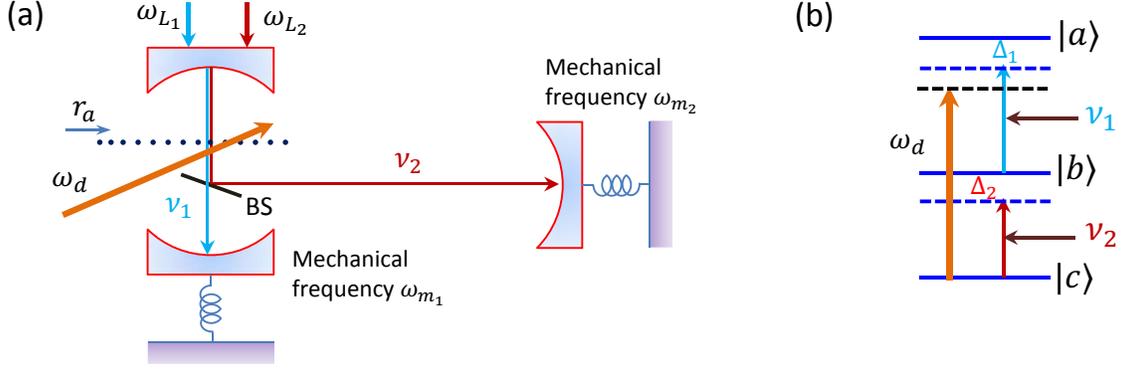}
\caption{(a) Schematics of a two-mode correlated spontaneous emission laser coupled to movable mirrors of mechanical frequencies $\omega_{m_{1}}$ and $\omega_{m_{2}}$. The doubly-resonant cavity is driven by two external lasers of frequency $\omega_{L_{1}}$ and $\omega_{L_{2}}$, and the cavity modes, filtered by a beam splitter (BS), are coupled to their respective movable mirrors. (b) The gain medium of the laser system is an ensemble of three-level atoms in a cascade configuration injected at a rate $r_{a}$ into the cavity in a coherent superposition of the upper $|a\rangle$ and lower $|c\rangle$ levels. An external laser drive of amplitude $\Omega$  and frequency $\omega_{\rm d}$ is also applied to generate two-photon coherence by coupling the upper $|a\rangle$ and lower $|c\rangle$ levels.}\label{fig1}
\end{figure*}

In this work, we consider a two-mode laser with the two movable mirrors of the doubly-resonant cavity  coupled to the cavity fields via radiation pressure. The laser system consists of a gain medium of three-level atoms in a cascade configuration. We rigorously derive a master equation for the two-mode laser coupled to thermal reservoirs, which generalizes previous results that are only valid for the case of driven coherence \cite{Zub13}. Using this master equation and the mirror-field interaction Hamiltonian we obtain Langevin equations, which are used to study the bistability and entanglement between the two movable mirrors. We show that, in contrast to the conventional bistability in single-mode optomechanics \cite{Tre96,Dor83,Asp13} that is shown to exist only when the cavity frequency is larger than the laser frequency, the mean photon numbers exhibit bistability for \textit{all values of detuning} due to the intermode coupling induced by the two-photon coherence. Additionally, the bistabilities show anomalous (``ribbon''-shaped) hysteresis for the circulation of the intracavity intensities for cavity-laser detuning opposite to the conventional bistability frequency range. These anomalous bistabilities are observed only if the rotating wave approximation is not made in the coupled Langevin equations. We also investigate the entanglement of the movable mirrors as a result of coupling to the laser system and find that the movable mirrors are strongly entangled in the adiabatic regime using realistic parameters. Interestingly, the entanglement persists for environmental temperature of the mechanical oscillators up to $12~\text{K}$, making our system a source for robust entanglement.

\section{Model and Hamiltonian}
We consider a two-mode three-level laser with two movable mirrors. The schematics of the laser system is shown Fig. \ref{fig1}a. The active medium is an ensemble of three-level atoms in a cascade configuration; see Fig. \ref{fig1}b. The atoms, initially prepared in coherent superposition of the upper $|a\rangle$ and lower $|c\rangle$ levels with no population in the intermediate level $|b\rangle$, are injected into the doubly-resonant cavity at a rate $r_{a}$  and removed after a time $\tau$, longer than the spontaneous emission time. During this time each atom nonresonantly interacts with the two-cavity modes of frequencies $\nu_{1}$ and $\nu_{2}$. Moreover, the upper and lower levels are driven by a strong laser field of amplitude $\Omega$ and frequency $\omega_{\rm d}$. We treat the movable mirrors as harmonic oscillators. The doubly-resonant cavity is driven by two additional coherent drives.

The total Hamiltonian of the system in the rotating wave and dipole approximations is given by ($\hbar=1$) \cite{Scu-book97}
\begin{align}\label{af}
H&=\sum_{j=a,b,c}\omega_{j}|j\rangle\langle j|+\sum_{j=1}^{2}\nu_{j}a_{j}^{\dag}a_{j}\notag\\
&+g_{1}(a_{1}|a\rangle\langle b|+a_{1}^{\dag}|b\rangle\langle a|)+g_{2}(a_{2}|b\rangle \langle c|+a_{2}^{\dag}|c\rangle \langle b|)\notag\\
&+i\frac{\Omega}{2}(e^{-i\omega_{d}t}|a\rangle\langle c|-\text{h.c})+i\sum_{j=1}^{2}(\varepsilon_{j}a_{j}^{\dag}e^{-i\omega_{\text{L}_{j}}t}-\text{h.c.})\notag\\
&+\sum_{j=1}^{2}[\omega_{m_{j}}b_{j}^{\dag}b_{j}+G_{j}a_{j}^{\dag}a_{j}(b_{j}+b_{j}^{\dag})],
\end{align}
where $\omega_{j}~(j=a,b,c)$ is the frequencies of the $j^{\rm th}$ atomic level and $g_{1}~(g_{2})$ is the coupling strength between the transition $|a\rangle\rightarrow |b\rangle$ ($|b\rangle\rightarrow |c\rangle$) and the cavity mode, $a_{j}~(a_{j}^{\dag})$ is the annihilation (creation) operator for the $j^{\rm th}$ cavity mode; $\omega_{m_{j}}$ are the mechanical frequencies, $b_{j}~(b_{j}^{\dag})$ are the annihilation (creation) operators for the mechanical modes, and $G_{j}=(\nu_{j}/L_{j})\sqrt{\hbar/m_{j}\omega_{\rm m_{j}}}$ is the optomechanical coupling strength with $L_{j}$ and $m_{j}$ being the length of the cavities and the mass of the movable mirrors, respectively. $|\varepsilon_{j}|=\sqrt{\kappa_{j}P_{j}/\hbar \omega_{\rm L_{j}}}$ are the amplitude of the lasers that drive the doubly-resonant cavity, with $\kappa_{j}$, $P_{j}$, and $\omega_{\rm L_{j}}$ being the damping rates of the cavities, the power, and the frequencies of the pump lasers, respectively. In Eq. \eqref{af}, the first line represents the free energy of the atom and the cavity modes and the terms in the second line describe the atom-cavity mode interactions. The first term in the third line describes the coupling of the levels $|a\rangle$ and $|c\rangle$ by a strong laser, while the second term represents the coupling of the external laser drives with the cavity modes. The first and second terms in the fourth line represent the free energy of the mechanical oscillators and the optomechanical couplings, respectively.

Using the fact that $|a\rangle\langle a|+|b\rangle\langle b|+|c\rangle\langle c|=1$, the free Hamiltonian for the atom and cavity modes can be written (dropping the constant $\hbar \omega_{c}$) as $H_{0}'\equiv (\omega_{a}-\omega_{c})|a\rangle\langle a|+(\omega_{b}-\omega_{c})|b\rangle\langle b|+\nu_{1}a_{1}^{\dag}a_{1}+\nu_{2}a_{2}^{\dag}a_{2}$. In view of this, the total Hamiltonian $H$ can be rearranged as $H=H_{0}+H_{I}$:
\begin{align}\label{H1}
H_{0}=(\tilde \nu_{1}+\tilde \nu_{2})|a\rangle \langle a|+\tilde\nu_{2}|b\rangle\langle b|+\tilde \nu_{1}a_{1}^{\dag}a_{1}+\tilde \nu_{2}a_{2}^{\dag}a_{2}
\end{align}
\begin{align}
H_{I}&=(\Delta_{1}+\Delta_{2})|a\rangle\langle a|+\Delta_{2}|b\rangle \langle b|+\delta\nu_{1}a_{1}^{\dag}a_{1}+\delta\nu_{2}a_{2}^{\dag}a_{2}\notag\\
&+g_{1}(a_{1}|a\rangle\langle b|+a_{1}^{\dag}|b\rangle\langle a|)+g_{2}(a_{2}|b\rangle \langle c|+a_{2}^{\dag}|c\rangle \langle b|)\notag\\
&+i\frac{\Omega}{2}(e^{-i\omega_{d}t}|a\rangle\langle c|-\text{h.c})+i\sum_{j=1}^{2}(\varepsilon_{j}a_{j}^{\dag}e^{-i\omega_{\text{L}_{j}}t}-\text{h.c.})\notag\\
&+\sum_{j=1}^{2}[\omega_{m_{j}}b_{j}^{\dag}b_{j}+G_{j}a_{j}^{\dag}a_{j}(b_{j}+b_{j}^{\dag})],
\end{align}
where $H_{0}=H_{0}'-(\tilde \nu_{1}+\tilde \nu_{2})|a\rangle \langle a|-\tilde\nu_{2}|b\rangle\langle b|-\delta\nu_{1}a_{1}^{\dag}a_{1}-\delta\nu_{2}a_{2}^{\dag}a_{2}$, $\Delta_{1}=\omega_{ab}-\tilde \nu_{1}$ and $\Delta_{2}=\omega_{bc}-\tilde \nu_{2}$ with $\omega_{ab}=\omega_{a}-\omega_{b}$ and $\omega_{bc}=\omega_{b}-\omega_{c}$ being the frequencies for $|a\rangle\rightarrow |b\rangle$ and $|b\rangle\rightarrow |c\rangle$ transitions, respectively. Here we have introduced the shifted cavity mode frequencies $\tilde \nu_{j}\equiv\nu_{j}-\delta\nu_{j}$; the shifts $\delta\nu_{j}$ will be defined later in Secs. V and VI. Now the interaction picture Hamiltonian can be derived using the unitary transformation $\mathcal{H}=e^{iH_{0}t}H_{I}e^{-iH_{0}t}=\mathcal{H}_{1}+\mathcal{H}_{2}$:
\begin{align}
\mathcal{H}_{1}&=(\Delta_{1}+\Delta_{2})|a\rangle\langle a|+\Delta_{2}|b\rangle \langle b|+i\frac{\Omega}{2}(|a\rangle\langle c|-|c\rangle\langle a|)\notag\\
&+g_{1}(a_{1}|a\rangle\langle b|+a_{1}^{\dag}|b\rangle\langle a|)+g_{2}(a_{2}|b\rangle \langle c|+a_{2}^{\dag}|c\rangle \langle b|)\label{V1}
\end{align}
\begin{align}
\mathcal{H}_{2}&=\sum_{j=1}^{2}\big[\omega_{m_{j}}b_{j}^{\dag}b_{j}+\delta\nu_{j}a^{\dag}_{j}
a_{j}+G_{j}a_{j}^{\dag}a_{j}(b_{j}+b_{j}^{\dag})\notag\\
&+i(\varepsilon_{j}a_{j}^{\dag}e^{i\delta_{j}t}-\varepsilon_{j}^{*}a_{j}e^{-i\delta_{j}t})\big],\label{V2}
\end{align}
where $\delta_{j}=\tilde\nu_{j}-\omega_{\text{L}_{j}}$ and we have assumed a two-photon resonance condition $\omega_{d}=\tilde\nu_{1}+\tilde\nu_{2}$. We represent all terms that involve the atomic state by $\mathcal{H}_{1}$, which will be used to derive the master equation for the laser system, and the rest of the terms by $\mathcal{H}_{2}$. This is because it will be convenient to obtain the reduced master equation for the cavity modes only by tracing out the atomic states. See the next section for details.

In this work, the main idea is to exploit the two-photon coherence induced by the laser system to increase the mirror-mirror entanglement. We show that even though the movable mirrors are not directly coupled, the two-photon coherence induces an effective coupling between the two mirrors mediated by the cavity. This coupling strength also depends on the number of photons in the cavity. In effect, it is possible to improve the entanglement by increasing the input laser power (see; Figs. 6 b and 8).

\section{Master equation for the two-mode laser}
We next derive the reduced master equation for the cavity fields using the Hamiltonian Eq. \eqref{V1}. While there are several approaches for deriving the master equation, we here employ the procedure outlined in \cite{Set11b,Scu-book97}.
Suppose that $\rho_{AR}(t,t_{j})$ represent the density operator at time $t$ for the
radiation plus a single atom in the cavity that is injected at an earlier time $t_{j}$. Since the atom
leaves the cavity after time $\tau$, it easy to see that
$t-\tau\leq t_{j}\leq t$. Thus, the unnormalized density operator for an ensemble of atoms in the cavity plus the two-mode field at time $t$ can be written as
\begin{equation}\label{m1}
 \rho_{AR}(t)=r_{a}\sum_{j} \rho_{AR}(t,t_{j})\Delta
t,
\end{equation}
where $r_{a}\Delta t$ is the total number of atoms
injected into the cavity in a small time interval $\Delta
t$. Note that $\rho_{AR}(t)$ is normalized to the total number of atoms. In the limit that $\Delta t\rightarrow 0$, we can approximate the summation by integration. Differentiating both sides of the resulting equation yields
\begin{equation}\label{m2}
\frac{d}{dt}
 \rho_{AR}(t)=r_{a}\frac{d}{dt}\int_{t-\tau}^{t}
 \rho_{AR}(t,t^{\prime})dt^{\prime}.
\end{equation}
In order to include the initial preparation of the atoms into the dynamics, we expand the right-hand side of \eqref{m2}
\begin{align}\label{m3}
\frac{d}{dt} \rho_{AR}(t)&=r_{a}\Big\{[ \rho_{AR}(t,t)- \rho_{AR}(t,t-\tau)]\notag\\
&+\int_{t-\tau}^{t}\frac{\partial}{\partial
t} \rho_{AR}(t,t^{\prime})dt^{\prime}\Big\}.
\end{align}
Here $\rho_{AR}(t,t)$ represents the density operator for an atom plus the cavity modes at time $t$ for an atom injected at an ``earlier time'' $t$.
Assuming atomic and cavity mode states are uncorrelated at the
instant the atom is injected into the cavity (Markov approximation), the density operator for each field-atom pair can be written as \cite{Ber88}
$\rho_{AR}(t,t)\equiv \rho_{R}(t) \rho_{A}{(0)}$,
where $\rho_{R}(t)$ is the cavity modes density operator and $\rho_{A}{(0)}$ is the initial density operator for each atom.
For simplicity, we further assume that the states of atomic and cavity modes are uncorrelated just after the atom is removed from the cavity, i.e., the cavity modes does not change appreciably because of the interaction with an atom (or even several atoms) during time $\tau$.   This allows us to write $
 \rho_{AR}(t,t-\tau)\equiv \rho_{R}(t)\rho_{A}(t,t-\tau),
$ where $ \rho_{A}(t,t-\tau)$ is the density operator at time $t$ for an atom injected at $t-\tau$. In the following, for simplicity of notation, we represent the density of operator for the cavity modes by $\rho$ by dropping $R$ in $\rho_{R}$ for brevity.

In this work, we consider the atoms to be injected into the cavity in a coherent superposition of the upper $|a\rangle$ and lower $|c\rangle$ levels, that is, $|\psi_{A}(0)\rangle=c_a|a\rangle+c_c|c\rangle$. The corresponding initial density matrix of the atom then has the form $
\rho_{A}(0)=|\psi_{A}\rangle\langle \psi_{A}|=\rho_{aa}^{(0)}|a\rangle\langle a|+\rho_{cc}^{(0)}|c\rangle \langle c|+(\rho_{ac}^{(0)}|a\rangle \langle c|+\text{h.c.}),
$ where $\rho_{aa}^{(0)}=|c_a|^2$ and $\rho_{cc}^{(0)}=|c_c|^2$ are the upper and lower levels initial populations and $\rho_{ac}^{(0)}=c_a^{*}c_c$ is the initial two-photon atomic coherence. Such a coherence has been shown to produce two-mode squeezing and entanglement between the cavity modes \cite{Zub05,Ale07a,Ale07c,Set08}. Here we exploit this coherence to generate entanglement between the movable mirrors instead.

Using the assumption that the atom and the cavity field state are uncorrelated at the time of injection and when the atom leaves the cavity, Eq. \eqref{m3} can be put in the form
\begin{equation}\label{m5}
\frac{d}{dt}\rho_{AR}(t)=r_{a}\Big\{[ \rho_{A}(0)- \rho_{A}(t-\tau)] \rho+\int_{t-\tau}^{t}\frac{\partial}{\partial
t}\rho_{AR}(t,t^{\prime})dt^{\prime}\Big\}.
\end{equation}

Furthermore, the time evolution of the density operator $\rho_{AR}(t,t^{\prime})$ has the usual form
$\partial\rho_{AR}(t,t')/\partial t$ $=-i[\mathcal{H}_{1}, \rho_{AR}(t,t')]$,
which together with $\partial\rho_{AR}(t)/\partial t$$=r_{a}\int_{t-\tau}^{t}(\partial \rho_{AR}(t,t')/\partial t)dt'$ leads to
\begin{equation}\label{m7}
\frac{d}{dt} \rho_{AR}(t)=r_{a}[\rho_{A}(0)-\rho_{A}(t-\tau)] \rho-i[
\mathcal{H}_{1}, \rho_{AR}(t)].
\end{equation}
We are interested in the dynamics of the cavity modes only. As such, we trace the atom plus field density operator over the atomic variables to find
\begin{equation}\label{m8}
\frac{d}{dt} \rho(t)=-i\text{Tr}_{A}\left[\mathcal{H}_{1},\rho_{AR}(t)\right],
\end{equation}
where we have used the fact that $\text{Tr}_{A}[\rho_{A}(0)]=\text{Tr}_{A}[\rho_{A}(t-\tau)]=1$.
Substituting the Hamiltonian $\mathcal{H}_{1}$ in Eq. \eqref{m8} and performing the trace operation, we obtain
\begin{align}\label{m9}
\frac{d}{dt} \rho(t)=&-ig_{1}(a_{1} \rho_{ba}- \rho_{ba}a_{1}+a_{1}^{\dag} \rho_{ab}- \rho_{ab}a^{\dag}_{1})\notag\\
&-ig_{2}( a_{2} \rho_{cb}- \rho_{cb} a_{2}+ a_{2}^{\dag}\rho_{bc}- \rho_{bc} a_{2}^{\dag})\notag\\
&+\kappa_{1}\mathcal{L}[a_{1}]\rho+\kappa_{2}\mathcal{L}[a_{2}]\rho.
\end{align}
The Lindblad dissipation terms \cite{Wal08} in the last line with $\kappa_{j}$ being the cavity damping rates are added to account for the damping of the cavity modes by thermal reservoirs. The explicit form of these terms will be given later [see Eq. \eqref{master}]. The next step in the derivation of the master equation is to obtain conditioned density operators, $\rho_{ab}=\langle a|\rho_{AR}|b\rangle$, $ \rho_{bc}=\langle b|\rho_{AR}|c\rangle$ and their complex conjugates that appear in Eq. \eqref{m9}. To this end, we return to Eq. \eqref{m7} and solve for these elements. Now multiplying Eq. \eqref{m7} on the left by $\langle l |$ and on the right by $|k\rangle$, where $l,k=a,b,c$ and assuming that the atom decays to energy levels other than the three lasing levels when it leaves the cavity, i.e., $\langle l|\rho_{A}(t-\tau)]|k\rangle=0$, we obtain
\begin{align}\label{m10}
\frac{d}{dt} \rho_{lk}(t)&=r_{a}\rho_{lk}^{(0)}\rho-i\langle l|[\mathcal{H}_{1}, \rho_{AR}(t)]|k\rangle-\gamma_{lk} \rho_{lk}.
\end{align}
We phenomenologically included the last term to account for the spontaneous emission and dephasing processes. $\gamma_{l}\equiv\gamma_{ll}$ are the atomic spontaneous emission rates and $\gamma_{lk}(l\neq k)$ are the dephasing rates. Thus, using Eq. \eqref{m10}, the equations for $\rho_{ab}$ and $\rho_{bc}$ are
\begin{eqnarray}
\dot{\rho}_{ab}=&-&(\gamma_{ab}+i\Delta_{1})\rho_{ab}+i g_{1}( \rho_{aa} a_{1}- a_{1} \rho_{bb})\notag\\
~~~~~~~~&+&i g_{2} \rho_{ac} a_{2}^{\dag}+\frac{\Omega}{2}  \rho_{cb},\label{m11}\\
\dot{ \rho}_{bc}=&-&(\gamma_{bc}+i\Delta_{2}) \rho_{bc}+i g_{2}( \rho_{bb} a_{2}- a_{2} \rho_{cc})\notag\\
&-&i g_{1}a_{1}^{\dag} \rho_{ac}-\frac{\Omega}{2} \rho_{ba}.\label{m12}
\end{eqnarray}
Here $\gamma_{ab}$ and $\gamma_{bc}$ are the dephasing rates for single-photon ``coherences'' $\rho_{ab}$ and $\rho_{bc}$, respectively.

To proceed further, we apply a linearization scheme, which amounts to keeping terms only up to second order in the coupling strength, $g_{j}$ in the master equation. This can be implemented by first writing the equations of motion for $ \rho_{aa}, \rho_{cc}, \rho_{ac}$, and $ \rho_{bb}$ to zeroth order in the coupling strength $g_{j}$ and substituting them in Eqs. \eqref{m11} and \eqref{m12} so that  $\rho_{ab}$ and $\rho_{bc}$ will be first order in $g_{j}$. Therefore, when the expressions for $\rho_{ab}$ and $\rho_{bc}$ are substituted in Eq. \eqref{m9}, the resulting master equation is second order in $g_{j}$.  Using Eq. \eqref{m1} the equations for $ \rho_{aa},  \rho_{cc}, \rho_{bb}$, and $ \rho_{ac}$ to first order in $g_{j}$ read
\begin{eqnarray}\label{m13}
  \dot { \rho}_{aa}&=&r_{a} \rho_{aa}^{(0)} \rho+\frac{\Omega}{2}(\rho_{ca}+ \rho_{ac})-\gamma_{a} \rho_{aa},~~~~\\
  \dot { \rho}_{cc} &=& r_{a}\rho_{cc}^{(0)} \rho-\frac{\Omega}{2} (\rho_{ac}+\rho_{ca})-\gamma_{c} \rho_{cc},~~~~\\
  \dot{ \rho}_{bb} &=&-\gamma_{b}\rho_{bb},\\
\dot {\rho}_{ac} &=& r_{a} \rho_{ac}^{(0)}\rho+\frac{\Omega}{2} (\rho_{cc}-\rho_{aa})-[\gamma_{ac}+i(\Delta_{1}+\Delta_{2})] \rho_{ac},~~~~\label{m14}
\end{eqnarray}
where $\gamma_{j}~(j=a,b,c)$ is the $j^{\rm th}$ atomic level spontaneous emission decay rates and $\gamma_{ac}$ is the two-photon dephasing rate. We next apply the good-cavity approximation, where the cavity damping rates $\kappa_{j}$ are much smaller than the spontaneous emission rates $\gamma_{j}$,  $\kappa_{j}\ll\gamma_{j}$. We also assume that $\kappa_{j}< r_{a}$. In this limit, the cavity modes vary more slowly than the atomic states, and thus the atomic states reach steady state in a short time. The time derivatives of such states can be set to zero while keeping the cavity-mode states time-dependent, which is frequently called the adiabatic approximation. After setting the time derivatives in Eqs. \eqref{m13}-\eqref{m14} to zero we obtain
\begin{align}\label{21}
&\rho_{aa}=\frac{r_{a}\rho}{d}Z_{aa},~~~\rho_{cc}=\frac{r_{a} \rho}{d}Z_{cc},\notag\\
&\rho_{ac}=\frac{r_{a} \rho}{d}Z_{ac},~~~\rho_{bb}=0,\notag
\end{align}
\begin{align}
 Z_{aa}&= \frac{1}{2}\{\gamma_{c}\chi(1-\eta)+\Omega^2\gamma_{ac}/2+ \gamma_{c}\gamma_{ac}\Omega\sqrt{1-\eta^2}\},~~~~~\notag\\
  Z_{cc}&=\frac{1}{2}\{\gamma_{a}\chi(1+\eta)+\Omega^2\gamma_{ac}/2+ \gamma_{a}\gamma_{ac}\Omega\sqrt{1-\eta^2} \},~~~~~\notag\\
Z_{ac}&=\frac{\sqrt{1-\eta^2}}{8[\gamma_{ac}+i(\Delta_1+\Delta_2)]}\Big\{4\mu-\Omega^2\gamma_{ac}(\gamma_{a}+\gamma_{c})\notag\\
  & -\frac{\chi\Omega}{4[\gamma_{ac}+i(\Delta_1+\Delta_2)]}[(1-\eta)\gamma_{b}-(1+\eta)\gamma_{a}],\notag
\end{align}
with $\chi = \gamma_{ac}^2+(\Delta_1+\Delta_2)^2$, $d = \gamma_a\gamma_c\chi+\Omega^2\gamma_{ac}(\gamma_{a}+\gamma_{c})/2$. In order to represent the initial state of the atoms with a single parameter, we have introduced a new variable $\eta\in[-1,1]$, such that the initial populations and coherence are given by $\rho_{aa}^{(0)}=(1-\eta)/2, \rho_{cc}^{(0)}=(1+\eta)/2$ and initial coherence $\rho_{ac}^{(0)}=\sqrt{1-\eta^2}/2$, respectively. Applying the adiabatic approximation in Eqs. \eqref{m11} and \eqref{m12} and using the expressions for $ \rho_{aa}, \rho_{bb}, \rho_{cc}$ and $\rho_{ac}$, we obtain after some lengthy algebra
\begin{eqnarray}
  -ig_{1} \rho_{ab}&=& \xi_{11}\rho a_{1}+\xi_{12} \rho a_2^{\dag},\label{mm1} \\
  ig_{2} \rho_{bc} &=& \xi_{22}a_{2}\rho+\xi_{21}a_{1}^{\dag}\rho, \label{mm2}
\end{eqnarray}
\begin{align}
\xi_{11}=\frac{g_{1}^2r_{a}}{\Upsilon d}[(\gamma_{bc}-i\Delta_{2})Z_{aa}+\frac{\Omega}{2} Z_{ac}^{*}],\\
\xi_{12}=\frac{g_{1}g_{2}r_{a}}{\Upsilon d}[(\gamma_{bc}-i\Delta_{2})Z_{ac}+\frac{\Omega}{2}Z_{cc}],\\
\xi_{21}=\frac{g_{1}g_{2}r_{a}}{\Upsilon^{*} d}[(\gamma_{ab}-i\Delta_{1})Z_{ac}-\frac{\Omega}{2} Z_{aa}],\\
\xi_{22}=\frac{g_{2}^2r_{a}}{\Upsilon^{*}d}[(\gamma_{ab}-i\Delta_{1})Z_{cc}-\frac{\Omega}{2} Z_{ac}^{*}],
\end{align}
where $\Upsilon=(\gamma_{ab}+i\Delta_{1})(\gamma_{bc}-i\Delta_{2})+\Omega^2/4$. Thus, substituting Eqs. \eqref{mm1} and \eqref{mm2} into Eq. \eqref{m9}, we obtain the master equation for just the cavity modes
\begin{align}\label{master}
\frac{d}{dt}\rho&=\xi_{11}(a_{1}^{\dag} \rho a_{1}-\rho a_{1}a_{1}^{\dag})+\xi_{11}^{*}(a_{1}^{\dag} \rho a_{1}- a_{1}a_{1}^{\dag} \rho)\notag\\
&+\xi_{22}(a_{2} \rho a_{2}^{\dag}-a_{2}^{\dag}a_{2} \rho)+\xi_{22}^{*}(a_{2}\rho a_{2}^{\dag}- \rho a_{2}^{\dag} a_{2})\notag\\
&+\xi_{12}(a_{1}^{\dag} \rho a_{2}^{\dag}-\rho a_{2}^{\dag}a_{1}^{\dag})+\xi_{12}^{*}(a_{2} \rho a_{1}- a_{1} a_{2} \rho)\notag\\
&+\xi_{21}(a_{1}^{\dag}\rho a_{2}^{\dag}- a_{2}^{\dag}a_{1}^{\dag}\rho)+\xi_{21}^{*}(a_{2}\rho a_{1}- \rho a_{1} a_{2})\notag\\
&+\frac{1}{2}\sum_{i=1}^{2}\kappa_{i}\Big[ (N_{i}+1)(2a_{i}\rho a^{\dag}_{i}-a^{\dag}_{i}a_{i} \rho-\hat \rho a^{\dag}_{i}a_{i})\notag\\
&+N_{i}(2a_{i}^{\dag}\rho a_{i}-a_{i}a^{\dag}_{i}\rho-\rho a_{i}a^{\dag}_{i})\Big].
\end{align}
Here we included the damping of the cavity modes by two independent thermal reservoirs with mean photon number $N_{j}$. Note that the terms proportional to $\text{Re}(\xi_{11})$ give rise to gain for the first cavity mode while $\text{Im}(\xi_{11})$ yields a frequency shift. The terms proportional to $\text{Re}(\xi_{22})$ result in loss of the second cavity mode while $\text{Im}(\xi_{22})$ produces a frequency shift. The terms proportional to $\xi_{12}$ and $\xi_{21}$ represent the correlation between the two cavity modes, which are known to produce two-mode squeezing and entanglement between the cavity modes \cite{Ale07a, Ale07c,Set08,Zub05}. In this work, we now exploit this correlation to entangle the movable mirrors of the doubly-resonant cavity.

\section{Quantum Langevin equations}
To analyze the bistability and entanglement between the two movable mirrors it is more convenient to use the quantum Langevin approach. In this respect, we derive the quantum Langevin equation for the atom-cavity mode and the optomechanical system separately. This is justified if the atom-field coupling is much stronger than the optomechanical coupling, which is the regime considered in this work. The contribution of the laser system (without mechanical oscillators) to the Langevin equations for the cavity field is derived from the master equation \eqref{master} using $\langle \dot o\rangle=\text{Tr}(\dot \rho o),~(o=a_{1}, a_{2})$  and removing the bracket from the resulting equations by adding appropriate noise operators $F_{j}$  with vanishing mean $\langle F_{j}\rangle=0$ \cite{Scu-book97}
\begin{align}
  \dot a_{1}=&-\frac{1}{2}(\kappa_{1}-2\xi_{11})a_{1}+\xi_{12}a_{2}^{\dag}+F_{1},\label{11}\\
\dot a_{2}=&-\frac{1}{2}(\kappa_{2}+2\xi_{22})a_{2}-\xi_{21}a_{1}^{\dag}+F_{2}.\label{22}
\end{align}
The correlation properties of the noise operators can be obtained by using Einstein relations \cite{Scu85}:$
  \langle D_{o_{1}o_{2}}\rangle=\frac{d}{dt}\langle o_{1}o_{2}\rangle-\langle(\dot o_{1}-F_{o_{1}})o_{2}\rangle-\langle o_{1}(\dot o_{2}-F_{o_{2}}),$
where $\langle  D_{o_{1}o_{2}}\rangle$ is the diffusion coefficient (with $o_{j}=a_{j},a_{j}^{\dag}$). Using this relation and the equations for second-order moments of the cavity modes operators $a_{j}$, the nonvanishing correlation properties of the noise operators are:
\begin{eqnarray}
\langle F_{1}^{\dag}(t)F_{1}(t')\rangle &=& [\kappa_{1}N_{1}+2\text{Re}(\xi_{11}) ]\delta(t-t'),\\
\langle F_{1}(t)F_{1}^{\dag}(t')\rangle &=& \kappa_{1}(N_{1}+1)\delta(t-t'),\\
\langle F_{2}^{\dag}(t)F_{2}(t')\rangle &=& \kappa_{2}N_{2}\delta(t-t'), \\
\langle F_{2}(t)F_{2}^{\dag}(t')\rangle &=& [\kappa_{2}(N_{2}+1)+2\text{Re}(\xi_{22})]\delta(t-t'), \\
\langle F_{2}(t)F_{1}(t')\rangle &=& -(\xi_{12}+\xi_{21})\delta(t-t').
  \end{eqnarray}
Now adding the contribution of the optomechanical coupling [Eq. \eqref{V2}] to the Langevin equations, we obtain the following equations for the cavity mode and mechanical mode operators:
\begin{align}
\dot a_{1}=&-(\frac{\kappa_{1}}{2}+i\delta\nu_{1}-\xi_{11})a_{1}+\xi_{12}a_{2}^{\dag}-iG_{1}a_{1}(b_{1}^{\dag}+b_{1})\notag\\
&+\varepsilon_{1}e^{i\delta_{1}t}+F_{1},\label{a1}\\
\dot a_{2}=&-(\frac{\kappa_{2}}{2}+i\delta\nu_{2}+\xi_{22})a_{2}-\xi_{21}a_{1}^{\dag}-iG_{2}a_{2}(b_{2}^{\dag}+b_{2})\notag\\
&+\varepsilon_{2}e^{i\delta_{2}t}+F_{2},\\
\dot b_{j}=&-i\omega_{\rm m_{j}}b_{j}-\frac{\gamma_{\rm m_{j}}}{2}b_{j}-i G_{j}a_{j}^{\dag}a_{j}+\sqrt{\gamma_{\rm m_{j}}}f_{j},\label{b}
\end{align}
where $f_{j}$ are the noise operators for the mechanical oscillators with zero mean and the following nonvanishing correlation properties:
\begin{align}
&\langle f_{j}^{\dag}(t)f_{j}(t')\rangle=n_{j}\delta(t-t'), \notag\\
&\langle f_{j}(t)f_{j}^{\dag}(t')\rangle=(n_{j}+1)\delta(t-t'),
\end{align}
where $n_{j}^{-1}=\exp(\hbar\omega_{\rm m_{j}}/k_{B}T_{j})-1$, $k_{B}$ is the Boltzmann constant and $T_{j}$ the temperature of the $j^{\rm th}$ thermal phonon bath. In the following sections, Eqs. \eqref{a1}-\eqref{b} will be used to study the bistability and entanglement between the two movable mirrors.

\section{Bistability of intracavity mean photon numbers}
Here we discuss the effect of the coupling induced by the two-photon coherence on the bistability of the mean intracavity photon numbers. It is well-known that the usual single-mode dispersive optomechanical coupling gives rise to an S-shaped bistability in the mean cavity photon number in the red-detuned frequency regime \cite{Asp13,Set12}. The bistability behaviour can be studied from the steady state solutions of the expectation values of  Eqs. \eqref{a1}-\eqref{b}. This can be done by first choosing a rotating frame defined by $\tilde a_{j}=a_{j}e^{-i\delta_{j}t}$ and by writing $\tilde a_{j}=\langle \tilde a_{j}\rangle+\delta \tilde a_{j}$ and $ b_{j}=\langle b_{j}\rangle+\delta b_{j}$. In this transformed frame, the equations for both the fluctuations $\delta \tilde a_{j}$ and classical mean values $\langle \tilde a_{j}\rangle$ have a coupling between the two cavity modes (terms proportional to $\xi_{12}$ and $\xi_{21}$) that contains highly oscillating factors $\exp[-i(\delta_{1}+\delta_{2})t]$. To obtain solutions for $\langle \tilde a_{j}\rangle$ in the steady state, one must either make rotating wave approximation, which amounts to dropping the highly oscillating terms completely or choose a condition such that $\delta_{2}=-\delta_{1}$ and retain the coupling terms. (It is important to mention here that we do not make the rotating wave approximation in the equations for the fluctuation $\delta \tilde a_{j}$, which later be used to study mirror-mirror entanglement.) In the following, we consider both cases and study the bistability of the intracavity photon numbers.

\textit{Rotating wave approximation (RWA)}. If we drop the highly oscillating terms (rotating wave approximation) in the transformed Langevin equations for $\langle \tilde a_{j}\rangle$, we obtain the steady state solutions for $\langle b_{j}\rangle$ and $\langle a_{j}\rangle$
\begin{align}
&\langle b_{j}^{\dag}+b_{j}\rangle=-\frac{2\omega_{\rm m_{j}} G_{j}I_{j}}{\gamma_{\rm m_{j}}^2/4+\omega_{\rm m_{j}}^2},\\
&\langle \tilde a_{j}\rangle=\frac{\varepsilon_{j}}{i\delta_{j}+(-1)^j\xi_{jj}+\kappa_{j}/2},\label{abs1}
\end{align}
where $I_{j}=|\langle \tilde a_{j}\rangle|^2$ are the steady state intracavity mean photon numbers and $\delta_{j} =\nu_{j}-\omega_{{L}_{j}}+G_{j}\langle b_{j}^{\dag}+b_{j}\rangle)$ are the cavity mode detuning. Here we have chosen $\delta\nu_{j}\equiv G_{j}\langle b_{j}^{\dag}+b_{j}\rangle$ to be  the frequency shift due to radiation pressure. The equations for the intracavity mean photon numbers have the implicit form
\begin{align}
I_{j}\left|i(\delta_{0j}-\beta_{j}I_{j})^2+\frac{\kappa_{j}}{2}+(-1)^j\xi_{jj}\right|^2=|\varepsilon_{j}|^2,\label{BRW}
\end{align}
where $\delta_{0j}=\nu_{j}-\omega_{{L}_{j}}$ and $\beta_{j}=(2\omega_{\rm m_{j}} G_{j}^2)/(\gamma_{\rm m_{j}}^2/4+\omega_{\rm m_{j}}^2)$. These are the standard equations for S-shaped bistabilities for intracavity intensities in an optomechanical system, with effective cavity damping rates $k_{j}+2(-1)^j\xi_{jj}$. Note that because of the RWA, there is no coupling between the intensities of the cavity modes that is due to the two-photon coherence induced in the system.
\begin{figure}
\includegraphics[width=4.2cm]{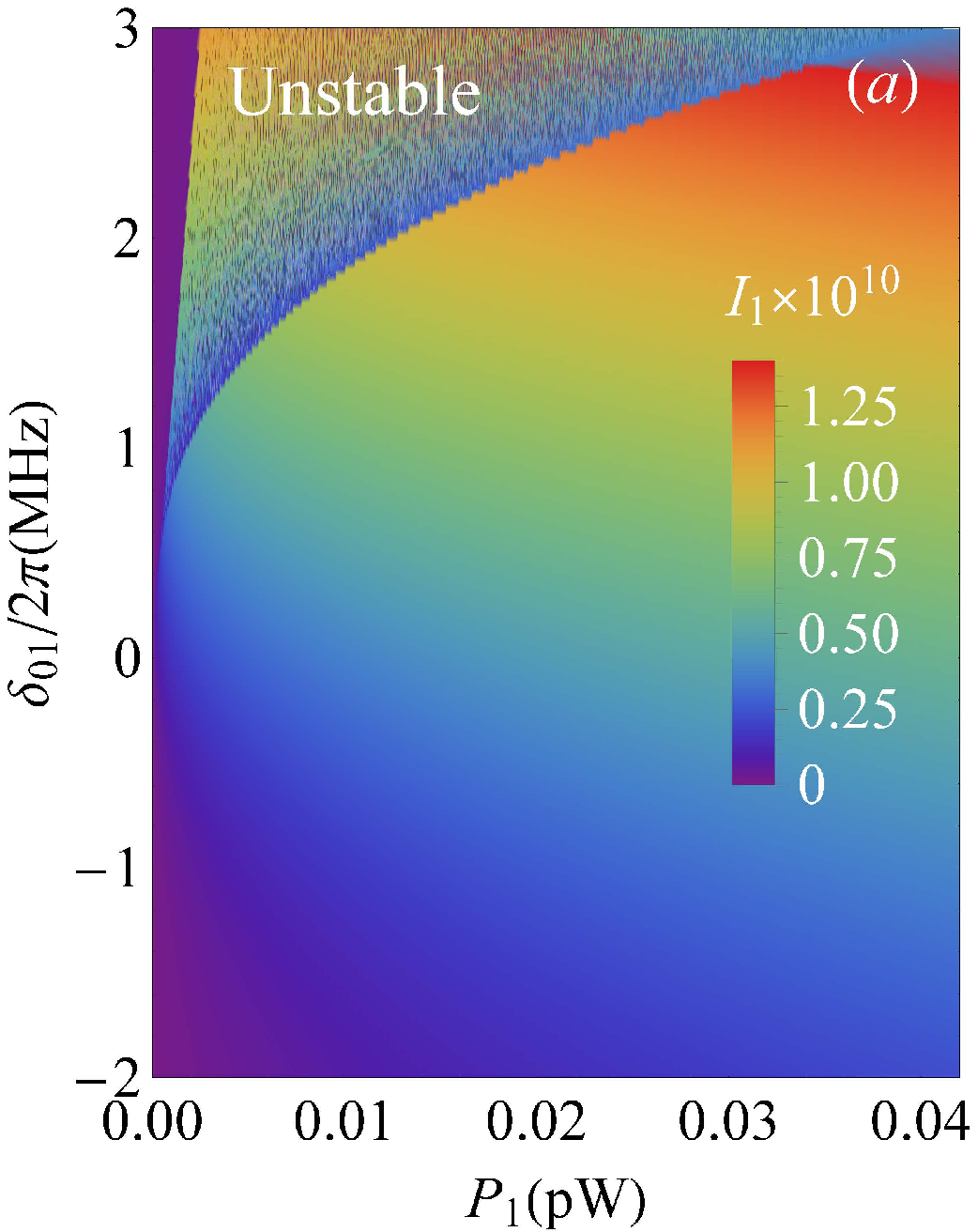}\includegraphics[width=4.2 cm]{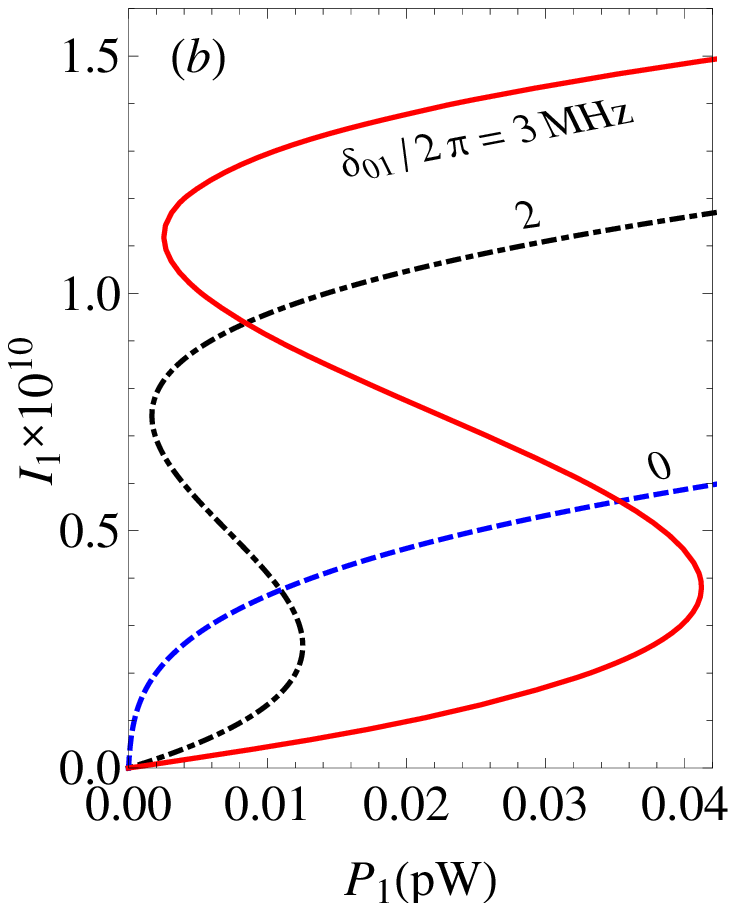}
\caption{(a) Phase diagram showing bistability of the intracavity mean photon number $I_{1}$ for varying cavity-laser detuning $\delta_{01}$ and cavity drive laser power $P_{1}$ in RWA. The meshed region shows the unstable solutions. (b) Cross section of the phase diagram for different values of the cavity-laser detuning $\delta_{01}$. Here we have used atom-field couplings $g_{1}=g_{2}=2\pi\times 4 ~\text{MHz}$, $\Omega/\gamma=10$ , $\kappa_{1}=\kappa_{2}=2\pi\times 215 ~\text{kHz}$ and when all atoms are initially in their excited state $|\psi_{0}\rangle=|a\rangle~(\eta=-1)$. See text for the other parameters.}\label{BS-2}
\end{figure}

Let us set realistic parameters from recent experiments \cite{Gro09,Arc06}: mass of the mirrors $m=145~ \text{ng}$, cavity lengths $L_{1}=112~ \mu\text{m}, L_{2}=88.6~\mu \text{m}$, pump laser wavelengths, $\lambda_{1}=810~\text{nm}$, $\lambda_{2}=1024\text{nm}$, rate of injection of atoms $r_{a}= 1.6~\text{MHz}$, mechanical oscillator damping rates $\gamma_{\rm m_{1}}=\gamma_{\rm m_{2}}=2\pi\times 60 ~\text{Hz}$, and mechanical frequencies $\omega_{\rm m_{1}}=\omega_{\rm m_{2}}=2\pi\times 3~\text{MHz}$, and dephasing and spontaneous emission rates for the atoms $\gamma_{ac}=\gamma_{ab}=\gamma_{bc}=\gamma_{a}=\gamma_{b}=\gamma_{c}=\gamma=3.4~\text{MHz}$. In this paper, we consider $\Delta_{1}=\Delta_{2}=0$ for the sake of simplicity.

To illustrate the bistability behaviour we plot, in Fig. \ref{BS-2}a, the steady state mean photon number for the first cavity mode $I_{1}$ as a function of the laser detuning and the cavity drive laser power $P_{1}$. This figure reveals that a large bistable regime (the meshed area) for a wide range of the drive laser power. As expected \cite{Asp13,Set12} the bistable behavior only exists for red-detuned ($\delta_{01}>0$) frequency range (notice that because of our definition of $\delta_{0j}=\nu_{j}-\omega_{{L}_{j}}$, red-detuned occurs for positive detuning which opposite to the usual convention \cite{Asp13}). The cross section of the phase diagram at different detunings shown in Fig. \ref{BS-2}b indicates the S-shaped bistable behaviour of the intracavity mean photon number $I_{1}$. We also observe that the bistable region widens with increasing detuning and drive laser power. Similar plots for the mean photon number $I_{2}$ show bistability for a wide range of detunings at a power one order of magnitude larger than that was needed to achieve the bistability of $I_{1}$, but we omit them here.

\begin{figure*}[t]
\includegraphics[width=5.5cm]{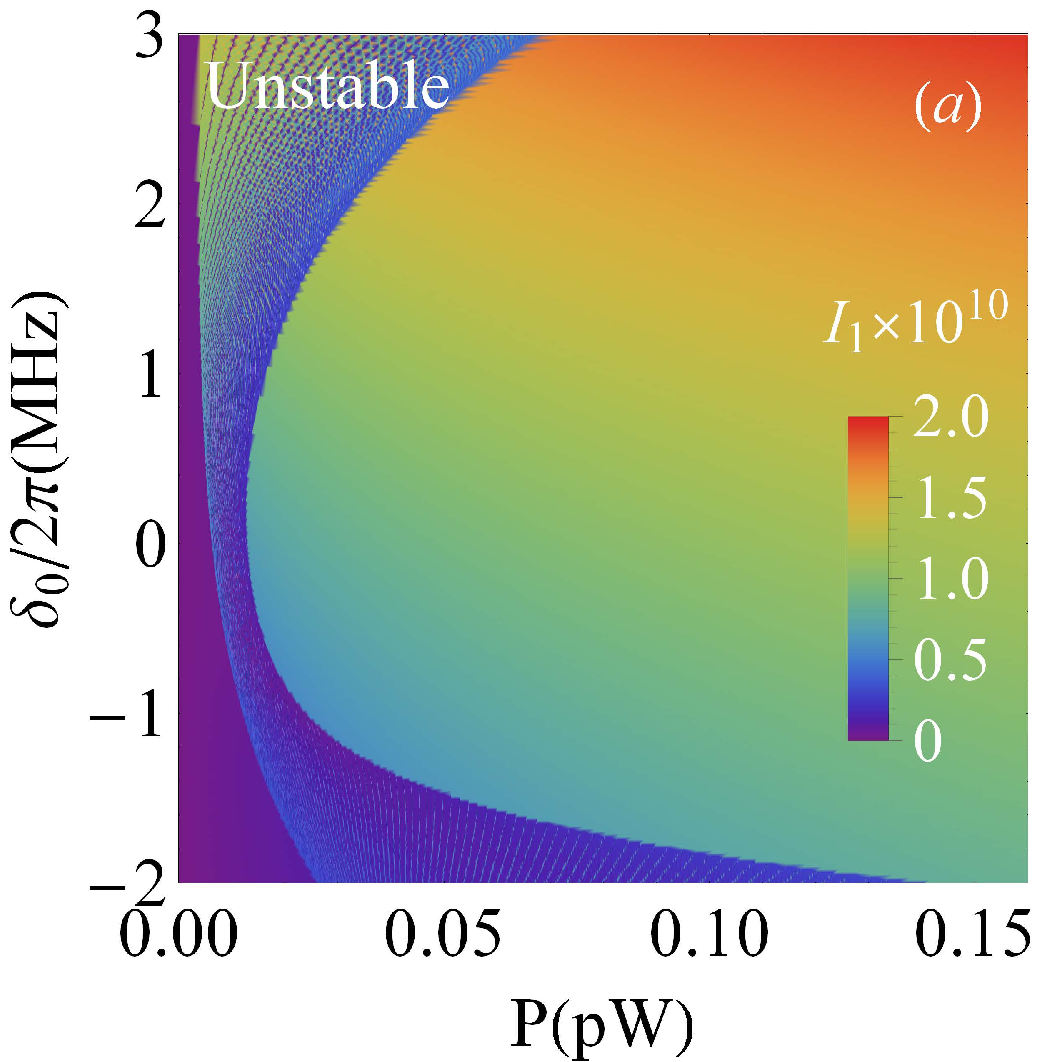} \includegraphics[width=5.5cm]{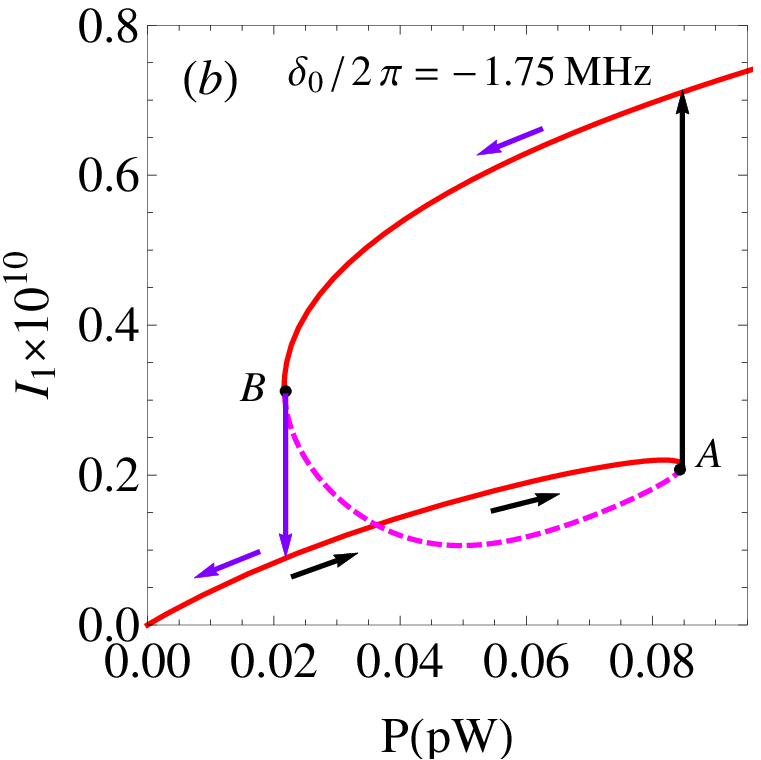} \includegraphics[width=5.5cm]{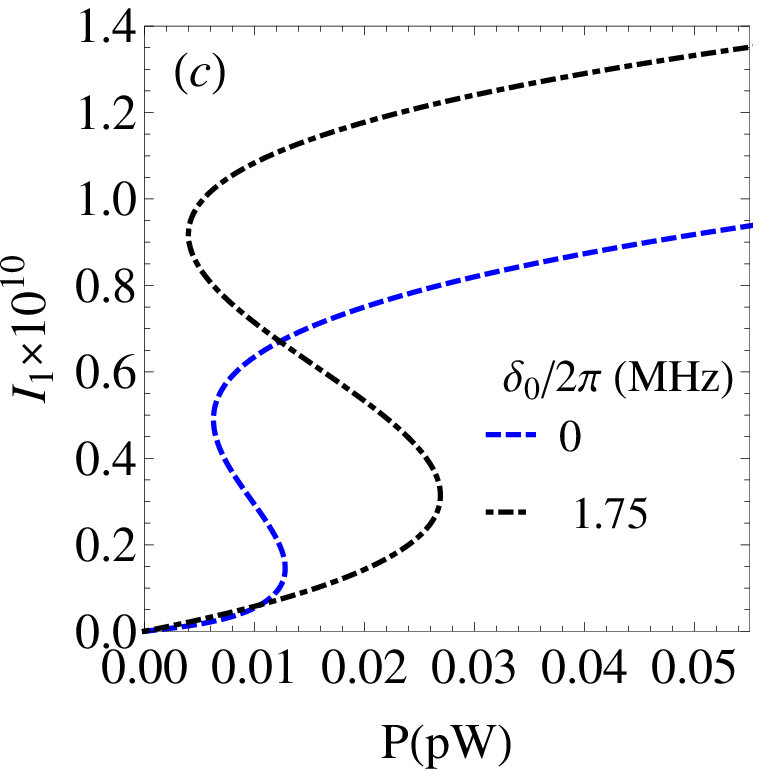}
\caption{(a) Phase diagram for mean photon number for the first cavity mode $I_{1}$ showing instability regions. The ``tornado''-shaped center area represents the unstable regime. Notice that the bistability appears for all values of detuning, which is in stark contrast to the usual red-detuned condition to observe bistability in single-mode optomechanics \cite{Set12,Asp13}. (b) Cross section of the phase diagram at $\delta_{0}/2\pi=-1.75~\text{MHz}$ showing anomalous ``ribbon''-shaped hysteresis due to the intermode coupling induced by the two-photon coherence. The area between the turning points represent the unstable regime, while the magenta-dashed curve show the saddle node instability. The arrows show the hysteresis for the circulation of the optical intensity. (c) Cross section of the phase diagram at $\delta_{0}=0$ (blue dashed curve) and $\delta_{0}/2\pi=1.75~\text{MHz}$ (black solid curve) showing the usual S-shaped bistability. Here we have used $\Omega/\gamma=10$, $\mu=0.1 ~(P_{2}=0.08 P_{1})$ and atoms are initially injected into the cavity in state $|c\rangle~(\eta=1)$. See text and Fig. \ref{BS-2} for the other parameters. }\label{fig3}
\end{figure*}

\textit{Beyond rotating wave approximation}. It is interesting to study the bistability behavior of the intracavity mean photon numbers in the nonrotating wave approximation, because it allows us to see the effect of the two-photon coherence. Note that to analyze the bistability in this regime, it is convenient to work in the rotating frame defined by the bare cavity frequencies $\nu_{j}$, which is equivalent to choosing $\delta \nu_{j}=0$ in the Hamiltonian given by Eq. \eqref{V2}. Thus, the condition for retaining the counter rotating terms in the Langevin equations for $\tilde a_{j}$ becomes $\delta_{02}=-\delta_{01}\equiv - \delta_{0}$. With this choice of detuning, we obtain the expectation values of the cavity mode operators:
\begin{align}
&\langle \tilde a_{1}\rangle=\frac{\varepsilon_{1}\alpha_{2}^{*}+\varepsilon_{2}\xi_{12}}{\alpha_{1}\alpha_{2}^{*}+\xi_{12}\xi_{21}^{*}}, ~~\label{aas1}\\
&\langle \tilde a_{2}\rangle=\frac{\varepsilon_{2}\alpha_{1}^{*}-\varepsilon_{1}\xi_{21}}{\alpha_{1}^{*}\alpha_{2}+\xi_{12}^{*}\xi_{21}},\label{as1}
\end{align}
where $\alpha_{1}=i(\delta_{0}-\beta_{1}I_{1})+\kappa_{1}/2-\xi_{11}$ and $\alpha_{2}=-i(\delta_{0}+\beta_{2}I_{2})+\kappa_{2}/2+\xi_{22}$. We see from Eqs. \eqref{aas1} and \eqref{as1} that the coupling between $\langle \tilde a_{1}\rangle$ and $\langle \tilde a_{2}\rangle$ is due to $\xi_{12}$ and $\xi_{21}$ which are proportional to the coherence induced either by the coupling of atomic levels by an external laser or by injecting the atoms in a coherent superposition of upper and lower levels. Introducing a new variable which relates the cavity drive amplitudes, $|\varepsilon_2|=\mu |\varepsilon_1|\equiv\mu |\varepsilon|~(P_{2}\sim \mu^2 P_{1})$, we obtain coupled equations for $I_{1}$ and $I_{2}$
\begin{align}
&\frac{|\alpha_{1}(I_{1})\alpha_{2}^{*}(I_{2})+\xi_{12}\xi_{21}^{*}|^2}{|\alpha_{2}^{*}(I_{2})+\mu\xi_{12}|^2} I_{1}=|\varepsilon|^2,\label{I11}\\
&\frac{|\alpha_{1}^{*}(I_{1})\alpha_{2}(I_{2})+\xi_{12}^{*}\xi_{21}|^2}{|\mu\alpha_{1}^{*}(I_{1})- \xi_{21}|^2} I_{2}=|\varepsilon|^2.\label{I22}
\end{align}

To gain insight into the effect of the coupling on the bistability behavior of the cavity modes, we slightly simplify the above equations by choosing the value of $\mu^2$. Let us first consider the case when $\mu^2\ll1 ~(P_{2}\ll P_{1})$. Thus, the denominator in Eq. \eqref{I22} can be approximated as $|\mu\alpha_{1}^{*}-\xi_{21}|^2\approx |\mu(-i\delta_{0}+\kappa_{1}/2-\xi_{11}^{*})-\xi_{21}|^2$ for $\mu ^2\beta_{1}I_{1}/|\xi_{21}|^2\ll1$. In this case, the ratio of Eqs. \eqref{I11} and \eqref{I22} yields a cubic equation for $I_{2}$: $I_{1}=I_{2}|\alpha_{2}^{*}(I_{2})+\mu\xi_{12}|^2/|\mu(-i\delta_{0}+\kappa_{1}/2-\xi_{11})-\xi_{21}|^2$. This equation reveals that $I_{2}$ can exhibit bistability when the intensity of the first cavity mode is varied. In Fig. \ref{fig3}a we plot a phase diagram showing steady state solutions for the first cavity mode mean photon number $I_{1}$. The ``tornado''-shaped center region represents the unstable solutions for positive detuning while the regions on the left and right areas represent stable solutions. In the vicinity of resonance ($\delta_{0}=0$) the unstable areas diminishes. The region of the unstable behavior widens when the detuning is increased further to large negative values. The intriguing aspect is that, in contrast to the RWA case, the bistability occurs at resonance as well as for the blue-detuned regime ($\delta_{0}<0$). Furthermore, these bistabilities occur at higher pump powers than the positive detunings. The cross section of the phase diagram at different detunings reveals two distinct features of the bistability. When $\delta_{0}>0$, the system exhibits the usual S-shaped bistability as discussed in the RWA case. However, when $\delta_{0}<0$ and above a critical detuning $\delta_{0}/2\pi\approx 1.1~\text{MHz}$, the system shows unconventional bistability showing ``ribbon''-shaped hysteresis--see in Fig. \ref{fig3}b. The circulation of the intensity shows peculiar behavior: when the drive laser power is swept to higher powers, the first turning point $A$ is reached at $P\approx 0.085~\text{pW}$ and the hysteresis then follows the upward arrow to the upper branch. When the laser power is decreased to lower values, the hysteresis reaches to the second turning point $B (P\approx 0.022 ~\text{pW})$ and the hysteresis follows the downward arrow to the lower branch.

\begin{figure*}[top]
\includegraphics[width=5.7cm]{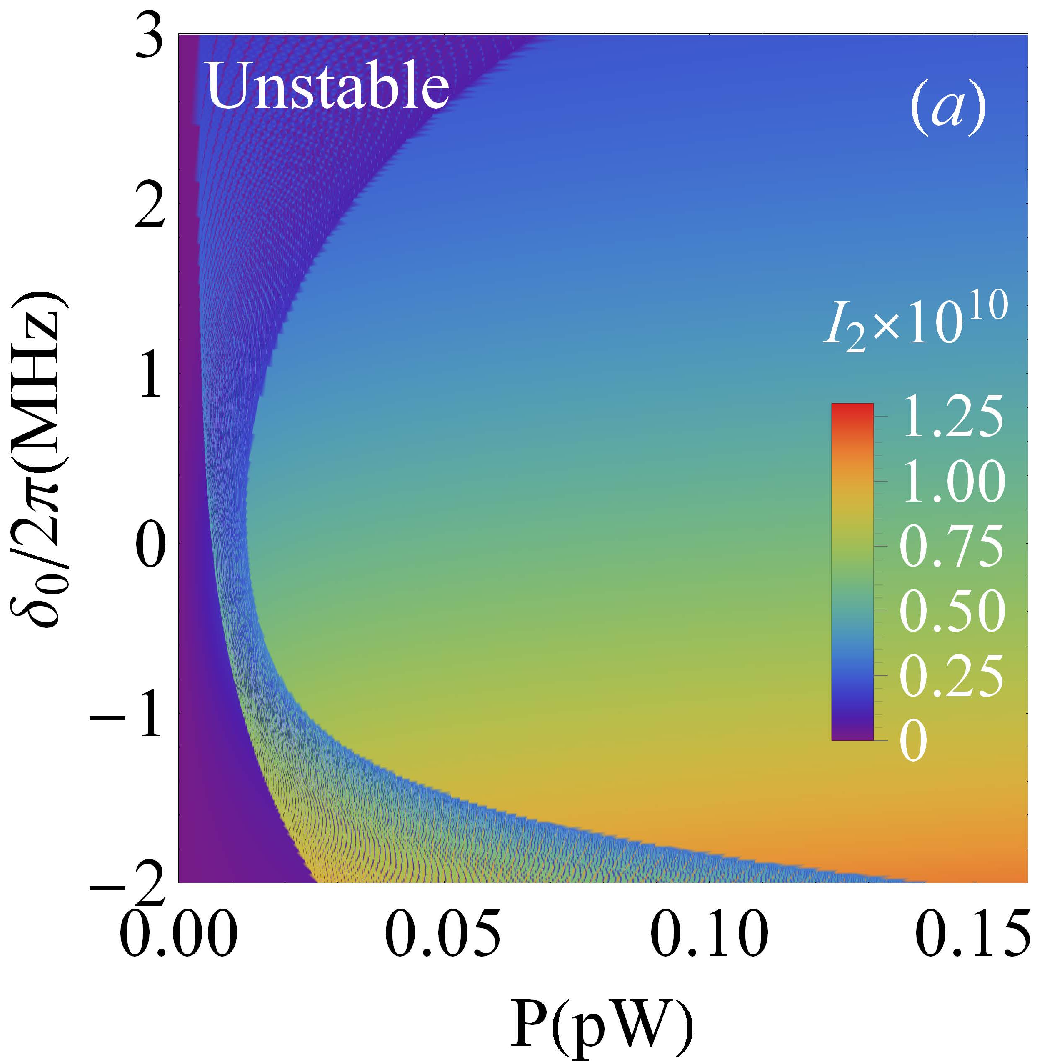}\includegraphics[width=5.9cm]{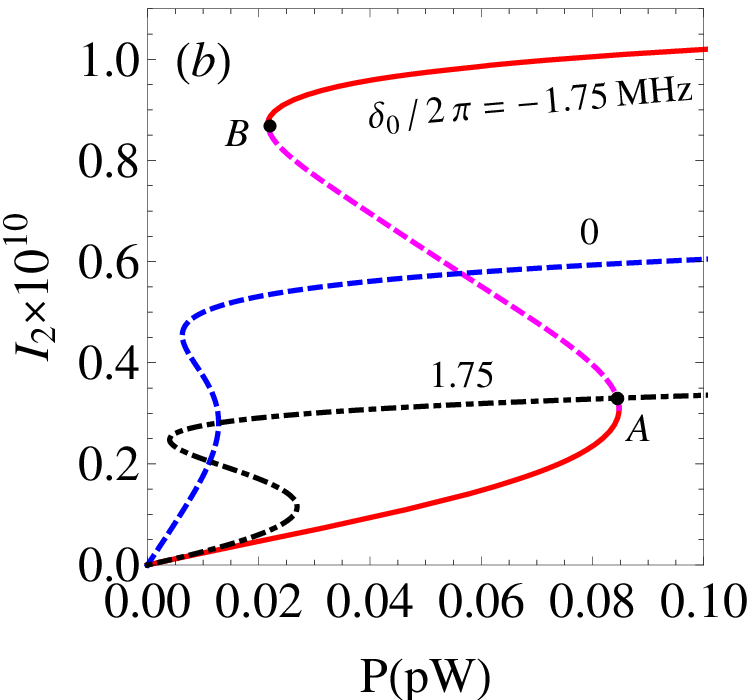}\includegraphics[width=5.7cm]{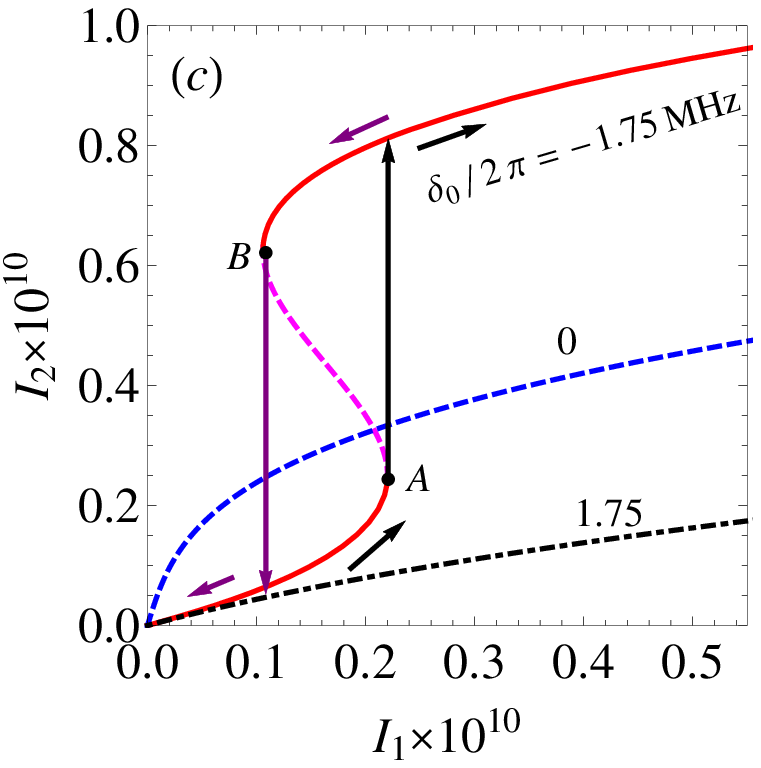}
\caption{(a) Phase diagram for mean photon number for the second cavity mode $I_{2}$ showing instability regions. The ``tornado''-shaped area represents the unstable regime. Notice that the bistability again appears for all values of detuning. (b) Cross section of the phase diagram at $\delta_{0}/2\pi=-1.75~\text{MHz}$ (red-solid curve) with a magenta-dashed curve showing the saddle node instability, $\delta_{0}=0$ (blue-dashed curve), and $\delta_{0}/2\pi=1.75~\text{MHz}$ (black-dotdashed curve) showing the usual S-shaped bistability. (c) Intracavity mean photon number for second mode $I_{2}$ vs the mean photon number for the first cavity mode $I_{1}$ indicating that $I_{2}$ exhibits S-shaped bistability behavior when $I_{1}$ is varied , only in the red-detuned ($\delta_{0}<0$) frequency range. The arrows indicate the hysteresis for the flow of intensities when $I_{1}$ is varied with turning points $A$ and $B$, which are the same turning points shown in Figs. \ref{fig3}b and \ref{fig4}b. The magenta-dashed curve shows the saddle node instability. The blue-dashed ($\delta_{0}=0$) and the black-dotdashed ($\delta_{0}/2\pi=1.75~\text{MHz}$) curves do not show bistability. Here we have used $\Omega/\gamma=10$, $\mu=0.1 ~(P_{2}=0.08 P_{1})$ and atoms are initially injected into the cavity in state $|c\rangle (\eta=1)$. See text and Fig. \ref{BS-2} for the other parameters.}\label{fig4}
\end{figure*}

In Fig. \ref{fig4}, we plot a phase diagram for the mean photon number of the second cavity mode $I_{2}$. Similar to the $I_{1}$, the mean photon number $I_{2}$ exhibits bistability for all values of detuning.  The main difference between the bistability behaviors of $I_{1}$ and $I_{2}$ is that $I_{2}$ only exhibits S-shaped bistability owing to the coupling between $I_{1}$ and $I_{2}$.  This can be understood from the bistability curve for $I_{2}$ when $I_{1}$ is varied. When $I_{1}$ increase from zero to higher values, $I_{2}$ also increases until a turning point $A$ (the same turning point shown in Fig. \ref{fig3}b and that of the red-solid  curve in Fig. \ref{fig4}b) is reached. The shape of the hysteresis for $I_{1}$ and $I_{2}$ is determined by whether the intensities increase or decrease along the saddle node instability curve (magenta-dashed curve in Fig. \ref{fig4}b). Notice that in traversing from turning points $A$ to $B$,  $I_{1}$ decreases but $I_{2}$ increases. Therefore, in the plot of $I_{1}$ versus power $P$ (see Fig. \ref{fig3}a), after the tuning point $A$, $I_{1}$ should decrease going below the turning point $A$ until the turning point $B$, producing the ``ribbon''-shaped bistability. However, since $I_{2}$ increases in going from $A$ to $B$, the saddle node instability curve in Fig. \ref{fig4}b should go above the turning point $A$ until it reaches $B$, creating the S-shaped bistability.

We next consider the case when $\mu^2\gg1~(P_{2}\gg P_{1})$. In this case, the denominator in Eq. \eqref{aas1} can be approximated as $|\alpha_{2}^{*}+\mu\xi_{12}|^2\approx|i\delta_{0}+\kappa_{2}/2+\xi_{22}+\mu\xi_{12}|^2$ assuming that $\beta_{2}I_{2}/(\mu^2|\xi_{12}|^2)\ll 1$. Then, the ratio of Eqs. \eqref{aas1} and \eqref{as1} gives a relation between $I_{1}$ and $I_{2}$: $I_{2}=I_{1}|\mu\alpha_{1}^{*}(I_{1})-\xi_{21}|^2/|i\delta_{0}+\kappa_{2}/2+\xi_{22}+\mu\xi_{12}|^2$. Therefore, $I_{1}$ can exhibit bistability behavior when $I_{2}$ is varied. Our numerical simulations (no shown here) reveals that, both $I_{1}$ and $I_{2}$ exhibit bistabilities for all values of detuning. However, for $\mu^2\gg1$, the role of $I_{1}$ and $I_{2}$ is interchanged: $I_{1}$ shows only S-shaped bistability while $I_{2}$ exhibits both S-shaped and unconventional bistability. In contrast to the case of $\mu^2\ll1$, the anomalous bistability emerges in the red-detuned ($\delta_{0}>0$) frequency range.

These rich features of intracavity mean photon number bistabilities are observed only if we do not make the rotating wave approximation in the steady state classical equations. This is because the rotating wave approximation drops the terms that couple the two cavity modes that are induced by the two-photon coherence, which is the main source of unconventional bistabilities. These unconventional bistabilities can be measured experimentally by measuring the field leaking out from the cavity. We expect that the transmitted field will also exhibit the bistability due to the linear input-output relation \cite{Wal08}.

\section{Entanglement of movable mirrors}
In this section we study the entanglement of the movable mirrors of the doubly-resonant cavity in the adiabatic regime. It has been shown that the cavity modes of the laser system are entangled \cite{Zub05,Ale07a,Ale07c,Set08} due to the two-photon coherence induced either by strong external drive or initial coherent superposition of atomic levels. Here we exploit this field-field entanglement to entangle the movable mirrors of the doubly-resonant cavity. Optimal entanglement transfer from the two-mode cavity field to the mechanical modes is achieved in the adiabatic limit, when the movable mirrors adiabatically follow the cavity fields, $\kappa_{j}\gg \gamma_{m_{j}}$ \cite{Pin05,Set14}, which is the case for mirrors with high-mechanical Q factor and weak effective optomechanical coupling.

Using the standard linearization procedure and transforming back (see Sec. V) to the original rotating frame by introducing $\delta a_{j}=\delta \tilde a_{j} e^{i \delta_{j}t}$ and defining $\tilde b_{j}=b_{j}\exp(i\omega_{m_{j}}t)$, we obtain
\begin{align}
\delta \dot a_{1}=&-\frac{\kappa_{1}'}{2}\delta a_{1}+\xi_{12}\delta a_{2}^{\dag}-iG_{1}\langle \tilde a_{1}\rangle(\delta b_{1}e^{-i(\omega_{m_{1}}-\delta_{1})t}\notag\\
&+\delta b_{1}^{\dag}e^{-i(\omega_{m_{1}}+\delta_{1})t})+F_{1}\\
\delta \dot a_{2}=&-\frac{\kappa_{2}'}{2}\delta a_{2}-\xi_{21}\delta a_{1}^{\dag}-iG_{2}\langle \tilde a_{2}\rangle(\delta b_{2}e^{-i(\omega_{m_{2}}-\delta_{2})t}\notag\\
&+\delta b_{2}^{\dag}e^{-i(\omega_{m_{2}}+\delta_{2})t})+F_{2}\\
\delta \dot {\tilde{b}}_{j}=&-\frac{\gamma _{m_{j}}}{2}\delta \tilde b_{j}-iG_{j}\langle \tilde a_{j}\rangle \delta
a_{j}^{\dag}e^{i(\omega_{m_{j}}+\delta_{j})t}\notag\\
&-iG_{j}\langle \tilde a_{j}^{\dag}\rangle \delta a_{j}e^{i(\omega_{m_{j}}-\delta_{j})t}+\sqrt{\gamma _{m_{j}}}f_{j},
\end{align}
where $\kappa_{1}'=\kappa_{1}-2\xi_{11}$ and $\kappa_{2}'=\kappa_{2}+2\xi_{22}$. Here $\langle \tilde a_{j}\rangle$ is given by Eq. \eqref{abs1}, which are obtained in RWA. We have deliberately made the rotating wave approximation to obtain the steady state solutions which would give stable solutions when choosing the effective detuning $\delta_{j}=\pm \omega_{\rm m_{j}}$. For $\delta_{j}=\pm \omega_{\rm m_{j}}$ the bistability of $I_{j}$ completely disappears, i.e., Eq. \eqref{abs1} becomes intensity independent. As mentioned earlier, no RWA has been made in the fluctuation equations  so that the coupling terms (proportional to $\xi_{12}$ and $\xi_{21}$) induced by the two-photon coherence are retained. In an optomechanical coupling when $\delta_{j}=\omega_{m_{j}}$, the interaction describes parametric amplification and can be used to generate optomechanical squeezing \cite{Asp13} and when $\delta_{j}=-\omega_{\rm m_{j}}$, the interaction is relevant for quantum state transfer \cite{Asp13,Pin05,Set14} and cooling.  Since we are interested in transferring the entanglement between the modes of the cavity to the mechanical modes, we choose $\delta_{j}=-\omega_{\rm m_{j}}$.

Setting $\delta_{j}=-\omega_{\rm m_{j}}$ and applying adiabatic approximation on the resulting $\delta a_{j}$ equations, we obtain coupled Langevin equations for $\tilde b_{j}$
\begin{align}
\delta \dot \tilde {b}_{1}&=-\frac{\Gamma_{1}}{2}\delta \tilde {b}_{1}-\mathcal{G}_{12}\delta b^{\dag}_{2}+v_1F_{1}+v_2F^{\dag}_{2}+\sqrt{\gamma_{m_{1}}}f_{1},\notag\\
\delta \dot \tilde {b}_{2}&=-\frac{\Gamma_{2}}{2}\delta \tilde {b}_{2}+\mathcal{G}_{21}\delta b^{\dag}_{1}-u_1F^{\dag}_{1}+u_2F_{2}+\sqrt{\gamma_{m_{2}}}f_{2},\notag
\end{align}
where $\Gamma_{j}=\gamma_{\rm m_{j}}+\Gamma_{b_{j}}$ with  $\Gamma_{b_{1}}=4\mathcal{G}_{1}^2\kappa_{2}'/K$ and $\Gamma_{b_{2}}=4\mathcal{G}_{2}^2\kappa_{1}'/K$ with $K=\kappa_{1}'\kappa_{2}'+4\xi_{12}\xi_{21}$ are the effective damping rate for the mechanical modes induced by the radiation pressure;  $\mathcal{G}_{12}=4\xi_{12} \mathcal{G}_{1}\mathcal{G}_{2}/K$ and $\mathcal{G}_{21}=4\xi_{21}\mathcal{G}_{1}\mathcal{G}_{2}/K$ are the effective coupling between the two mechanical modes induced by the laser system and $v_{1}=\sqrt{\Gamma_{b_{1}}\kappa_{2}'/K}$, $v_{2}=2\xi_{12}\sqrt{\Gamma_{b_{1}}/\kappa_{1}'K}$, $u_{1}=2\xi_{21}\sqrt{\Gamma_{b_{2}}/\kappa_{1}'K}$, and $u_{2}=\sqrt{\Gamma_{b_{2}}\kappa_{1}'/K}$. Here we have introduced many-photon coupling $\mathcal{G}_{j}=G_{j}\sqrt{|\langle \tilde a_{j}\rangle|}\equiv G_{j}\sqrt{I_{j}}$ by choosing the phase of the cavity laser drives such that $\langle \tilde a\rangle=-i|\langle \tilde a\rangle|$ \cite{Set14}. Note that since we have chosen $\Delta_{1}=\Delta_{2}=0$, for the sake of simplicity, $\xi_{jj}$ and $\xi_{ij}$ are real.

To analyze the entanglement between the two mechanical modes, it is convenient to use quadrature operators defined as $\delta q_{j}=(\delta \tilde {b}_{j}+\delta \tilde {b}_{j}^{\dag})/\sqrt{2}$, $\delta p_{j}=i(\delta \tilde {b}_{j}^{\dag}-\delta \tilde {b}_{j})/\sqrt{2}$. We also introduce the corresponding noise operators $f_{q_{i}}$, $f_{p_{i}}$ and $F_{x_{i}},F_{y_{i}}$ defined in a similar way. The equations for the these quadrature operators are
\begin{align}
\delta \dot q_{1}& =-\frac{\Gamma _{1}}{2}\delta q_{1}-\mathcal{G}_{12}\delta
q_{2}+\tilde{F}_{q_{1}},\\
\delta \dot{p}_{1}& =-\frac{\Gamma _{1}}{2}\delta p_{1}+\mathcal{G}_{12}\delta p_{2}+\tilde{F}_{p_{1}}, \\
\delta \dot q_{2}& =-\frac{\Gamma _{2}}{2}\delta q_{2}+\mathcal{G}_{21}\delta
q_{1}+\tilde{F}_{q_{2}},\\
\delta \dot{p}_{2}& =-\frac{\Gamma _{2}}{2}\delta p_{2}-\mathcal{G}_{21}\delta p_{1}+\tilde{F}_{p_{2}},
\end{align}
where $\tilde{F}_{q_{1}} =v_{2}F_{p_{2}}+v_{1}F_{q_{1}}+\sqrt{\gamma _{m_{1}}}
f_{q_{1}}$ ,
$\tilde{F}_{p_{1}} =-v_{2}F_{p_{2}}+v_{1}F_{p_{1}}+\sqrt{\gamma _{m_{1}}}
f_{p_{1}}$ ,
$\tilde{F}_{q_{2}} =u_{2}F_{q_{2}}-u_{1}F_{q_{1}}+\sqrt{\gamma _{m_{2}}}
f_{q2}$, and
$\tilde{F}_{p_{2}} =u_{2}F_{p_{2}}+u_{1}F_{p_{1}}+\sqrt{\gamma _{m_{2}}}
f_{p2}$. Alternatively, the above equations can be written in a matrix form as
\be\label{ME}
\dot U(t)=R U(t) +\zeta(t),
\ee
\begin{equation}\label{R}
R=\left(
\begin{array}{cccc}
-\Gamma _{1}/2 & 0 & -\mathcal{G}_{12} & 0 \\
0 & -\Gamma _{1}/2 & 0 & \mathcal{G}_{12} \\
\mathcal{G}_{21} & 0 & -\Gamma _{2}/2 & 0 \\
0 & -\mathcal{G}_{21} & 0 & -\Gamma _{2}/2%
\end{array}%
\right)
\end{equation}
and $U(t)=(\delta q_1,\delta p_1,\delta q_2,\delta p_2)^{T}$ and $\zeta(t)=(\tilde{F}_{q_{1}},\tilde{F}_{p_{1}},\tilde{F}_{q_{2}},\tilde{F}%
_{p_{2}})^{T}$.

\begin{figure}
\includegraphics[width=7cm]{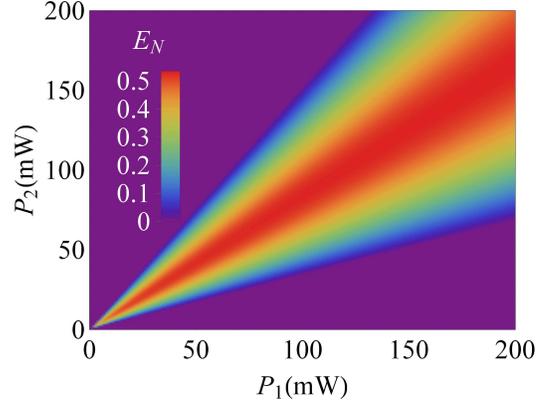}
\caption{ Entanglement of movable mirrors. Logarithmic negativity $E_{N}$ vs the cavity drive lasers' powers $P_{1}$ and $P_{2}$ for thermal phonon numbers $n_{1}=n_{2}=100$ and thermal photon numbers $N_{1}=N_{2}=1$, normalized drive laser amplitude $\Omega/\gamma=6$, $\eta=-1$ (more atoms are injected in their upper level $|a\rangle$), and atom-field coupling constants $g_{1}=g_{2}=2\pi\times 2.5 ~\text{MHz}$, and cavity damping rates $\kappa_{1}=2\pi\times 215 ~\text{kHz}$, and $\kappa_{2}=2\pi\times 430 ~\text{kHz}$. See text and Fig. \ref{fig5} for the other parameters.}\label{fig5}
\end{figure}

In this section, we focus on the steady state entanglement between the mechanical modes. To this end, one needs to find a stable solution for Eq. \eqref{ME} so that it reaches a unique steady state independent of the initial conditions. Since we have assumed the quantum noises $f_{q_{j}}, f_{p_{j}}$, $F_{x_{j}}$, and $F_{y_{j}}$ to be zero-mean Gaussian noises and the equations for fluctuations $(\delta q_j, \delta p_j$) are linearized, the quantum steady state for fluctuations is simply a zero-mean Gaussian state, which is fully characterized by a correlation matrix $V_{ij}=[\langle U_{i}(\infty)U_{j}(\infty)+U_{j}(\infty)U_{i}(\infty)\rangle]/2$. For fixed realistic parameters mentioned in this section, we have chosen externally controllable parameters such as $\Omega$, the powers of the cavity drive lasers, and initial state of the atoms for which the system is stable. Thus, for all results presented in this section the system is stable and the correlation matrix satisfies the Lyapunov equation
\begin{equation}\label{cm}
  RV+VR^{\rm T}=-D,
\end{equation}
\begin{equation}
D=\left(
\begin{array}{cccc}
A_{1} & 0 & A_{3} & 0 \\
0 & A_{1} & 0 & -A_{3} \\
A_{3} & 0 & A_{2} & 0 \\
0 & -A_{3} & 0 & A_{2}%
\end{array}%
\right)
\end{equation}
where $A_{1}=\kappa _{11}v_{1}^{2}+\kappa _{22}v_{2}^{2}-2\beta _{12}v_{1}v_{2}+%
\gamma _{m_{1}}(2n_{1}+1)$, $A_{3}=\beta _{12}(u_{1}v_{2}-u_{2}v_{1})+\kappa _{22}u_{2}v_{2}-\kappa _{22}u_{1}v_{1}$,
$A_{2}=\kappa _{11}u_{1}^{2}+\kappa _{22}u_{2}^{2}+2\beta _{12}u_{1}u_{2}+%
\gamma _{m_{2}}(2n_{2}+1)/2$ with $\kappa_{jj}\equiv[\kappa_{j}(2N_j+1)+2\text{Re}(\xi_{jj})]/2$ and $\beta_{12}\equiv\text{Re}(\xi_{12}+\xi_{21})/2$.
\begin{figure}
\includegraphics[width=4.5cm]{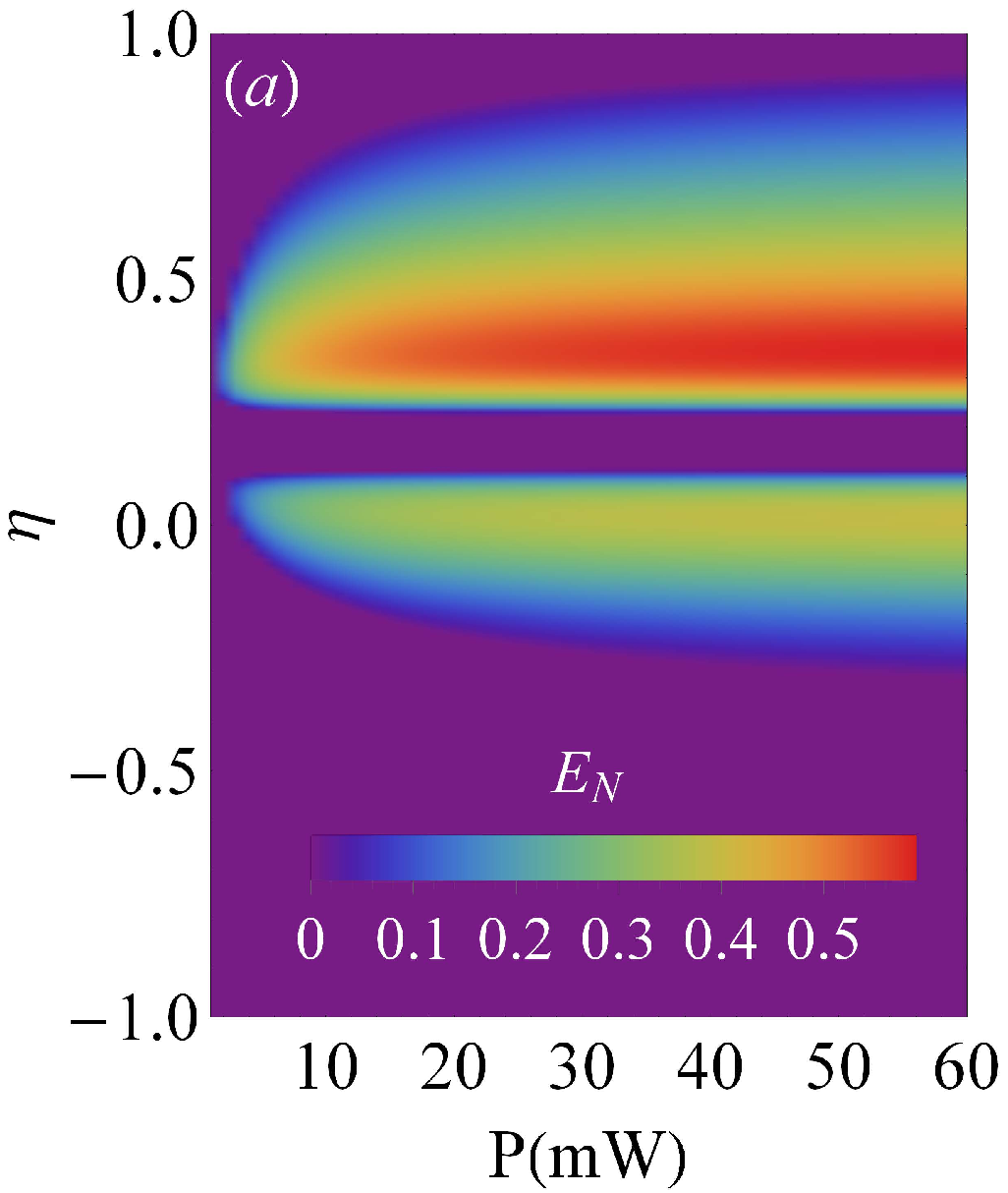}\includegraphics[width=4.2cm]{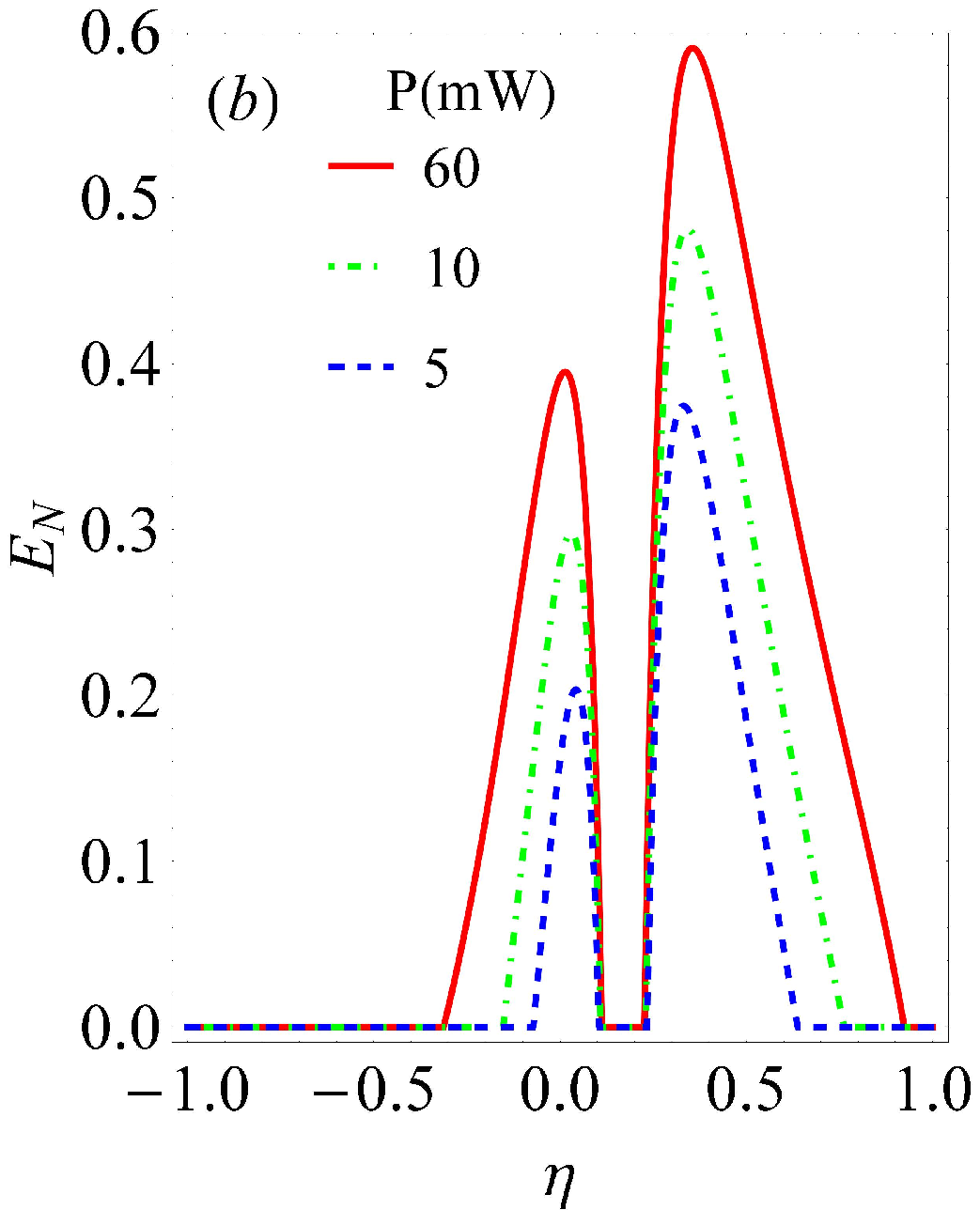}
\caption{ Entanglement of movable mirrors with injected coherence only ($\Omega=0$). Logarithmic negativity $E_{N}$ vs the cavity drive lasers power $P$ and initial state of the atoms $\eta$. For thermal phonon numbers $n_{1}=n_{2}=100$ and thermal photon numbers $N_{1}=N_{2}=1$. See text and Fig. \ref{fig5} for the other parameters.}\label{fig6}
\end{figure}

In order to quantify the two-mode entanglement, we employ the logarithmic negativity $E_{N}$, a quantity which has been proposed as a measure of bipartite entanglement for Gaussian states \cite{Vid02}. For continuous variables, $E_{N}$ is defined as
\begin{equation}\label{LN}
  E_{N}=\max [0,-\ln 2\Lambda],
\end{equation}
where $\Lambda=2^{-1/2}\left[\sigma-\sqrt{\sigma^2-4\text{det} V}\right]^{1/2}$ is the smallest simplistic eigenvalue of the partial transpose of the $4\times 4$ correlation matrix $V$ with $\sigma=\det V_{A}+\det V_{B}-2\det V_{AB}$. Here $V_{A}$ and $V_{B}$, respectively represent the first and second  mechanical modes, while $V_{AB}$ describes the correlation between them.  These matrices are elements of the $2\times2$ block form of the correlation matrix
\begin{equation}\label{vv}
  V\equiv \left(
                      \begin{array}{cc}
                        V_{A} & V_{AB}\\
                       V_{AB}^{T} & V_{B} \\
                      \end{array}
                    \right).
\end{equation}
The movable mirrors are entangled when the logarithmic negativity $E_{N}$ is positive.

In Fig. \ref{fig5} we plot the logarithmic negativity $E_{N}$ vs the cavity drive lasers' powers $P_{1}$ and $P_{2}$ when all atoms are injected in their upper level $|a\rangle$ ($\eta=-1$), for thermal phonon numbers $n_{1}=n_{2}=100$, and thermal photon numbers $N_{1}=N_{2}=1$. The two movable mirrors are entangled for a wide range of the drive lasers' powers. Maximum entanglement is achieved slightly below the diagonal of the phase diagram, i.e., when the drive laser power $P_{1}$ is slightly higher than $P_{2}$. This can be explained by the fact that the effective couplings $\mathcal{G}_{12}$ and $\mathcal{G}_{21}$ between the two mechanical mirrors can be enhanced because they directly rely on the mean number of photons $I_{j}$, or the cavity drive lasers' powers.

\begin{figure}
\includegraphics[width=4.6cm]{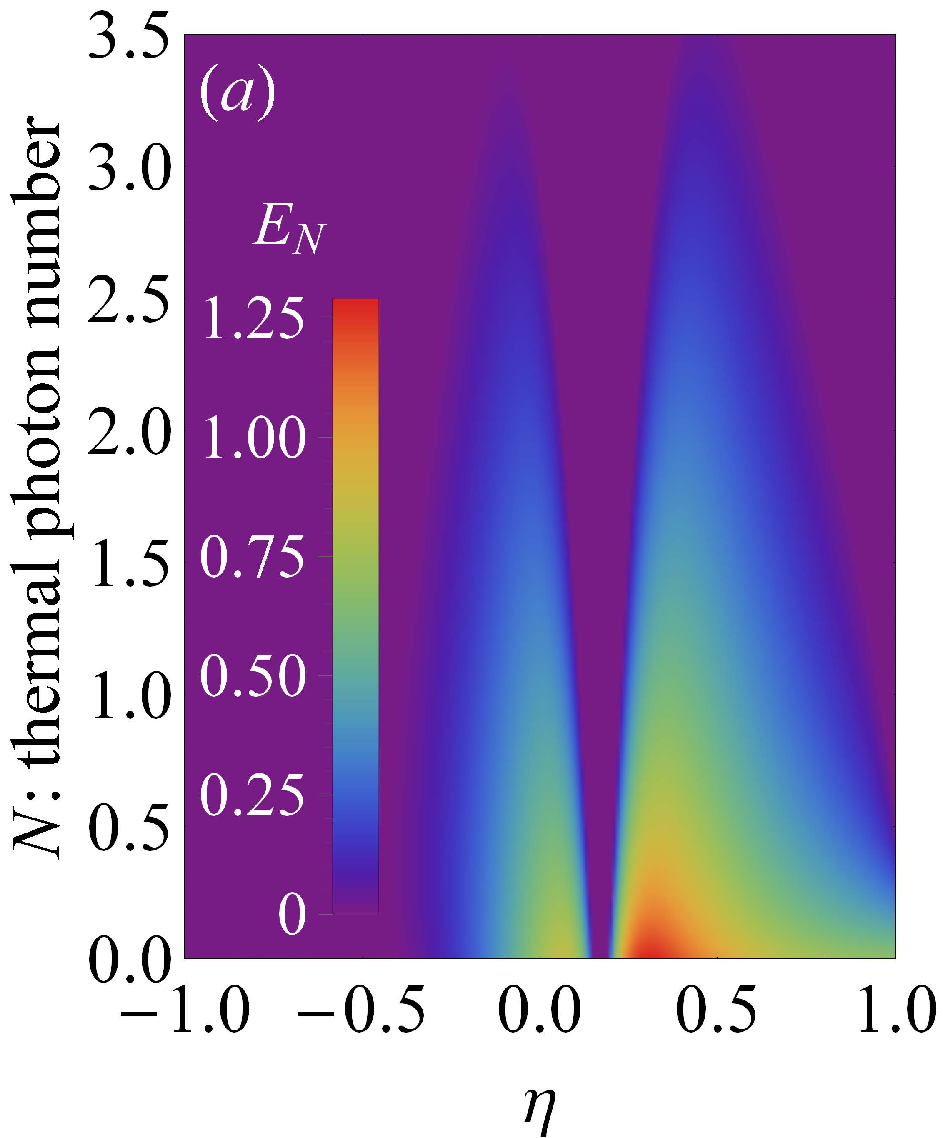}\includegraphics[width=4cm]{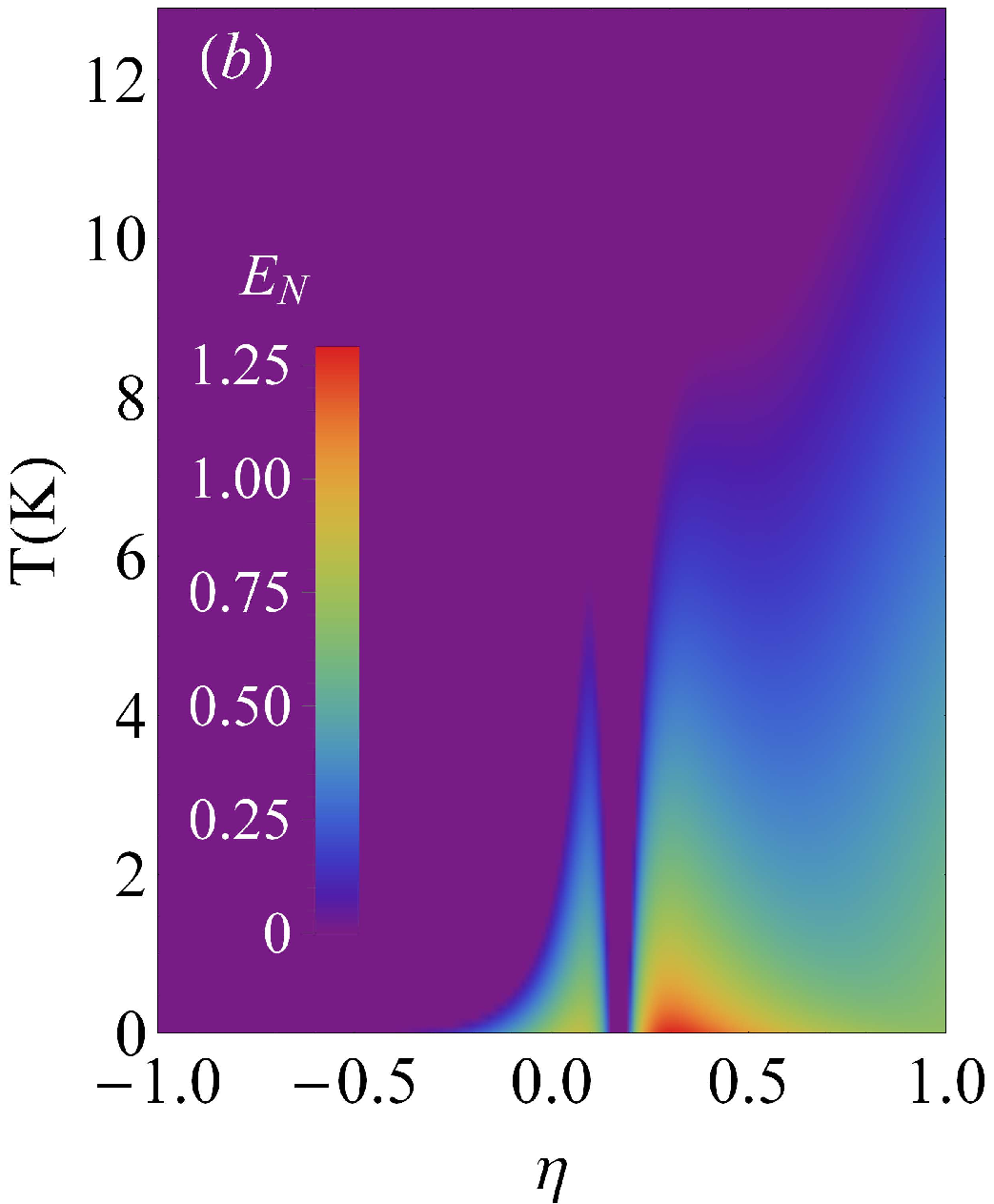}
\caption{Environment temperature dependence of the mirror-mirror entanglement with injected coherence only ($\Omega=0$). (a) Logarithmic negativity $E_{N}$ vs initial state of the atoms $\eta$ and thermal photon numbers $N$ when the temperature of the thermal phonon bath is zero $T=0~\text{K}~(n_{1}=n_{2}=0)$. (b) Logarithmic negativity $E_{N}$ vs the initial state of the atoms $\eta$ and the temperature $T$ of thermal phonon bath when the thermal photon bath is at zero temperature ($N=N_{1}=N_{2}=0$). The cavity drive lasers power is fixed at $P=P_{1}=P_{2}=200~\text{mW}$. See text and Fig. \ref{fig5} for the other parameters.}\label{fig66}
\end{figure}

We next examine the entanglement generated by either the driven or injected coherence separately. First, we consider the contribution of the injected coherence characterized by the initial states of the three-level atoms, i.e., $\eta$ to the entanglement of the mirrors. Figure \ref{fig6}a displays the phase diagram of logarithmic negativity as function of the cavity drive lasers power $P$ (assumed to be the same for both laser drives) and $\eta$. This figure reveals two blocks of parametric regimes showing entanglement of the two movable mirrors. The lower block appears around the maximum initial coherence $\eta=0$ [corresponds to  $|\psi_{A}(0)\rangle=(|a\rangle+|c\rangle)/\sqrt{2})$], while the second block appears for $\eta>0$, which corresponds to more atoms in the lower level than the upper level. It is somewhat counterintuitive that the maximum entanglement does not occur when the injected coherence is maximum. Instead, the maximum mirror-mirror entanglement is achieved around $\eta=0.36$, which corresponds to more atoms populating the upper level.

Figures \ref{fig66}a and \ref{fig66}b show the dependence of the entanglement on the temperature of the environment. When the cavity drive lasers' power is fixed at $P=200~\text{mW}$ and the temperature of the thermal phonon bath is zero $T=0\text{K}~ (n_{1}=n_{2}=0)$, the mirrors become disentangled at $N\approx3.5$. The range of $N$ for which the entanglement exists is weakly dependent on the drive power strength. However, when the thermal photon bath is at zero temperature $N=N_{1}=N_{2}=0$, the mirror-mirror entanglement persists up to a temperature $T\lesssim 12~\text{K}$ of the thermal phonon bath, which is two orders of magnitude larger than the ground state temperature of the movable mirrors. The entanglement can even survive at higher temperature if the drive laser power is increased. It is worth mentioning that the entanglement generated when more atoms are initially in the lower level ($\eta\gtrsim 0.3$) is more robust than that created around the maximum coherence $\eta\sim 0$. Therefore, the entanglement is robust against the thermal phonons temperature, but substantially more sensitive to the thermal photons temperature.

\begin{figure}
\includegraphics[width=7cm]{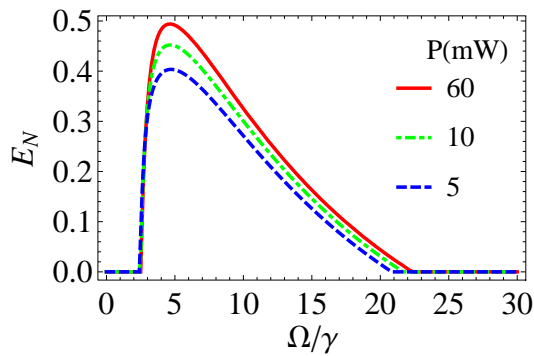}
\caption{Entanglement of movable mirrors with driven coherence only. Logarithmic negativity $E_{N}$ vs the cavity drive lasers power $P$ and normalized drive amplitude $\Omega/\gamma$ for thermal phonon numbers $n_{1}=n_{2}=100$ and thermal photon numbers $N_{1}=N_{2}=1$, in the absence of injected coherence $\eta=-1$ (all atoms are injected in the their upper level). See text and Fig. \ref{fig5} for the other parameters.}\label{fig8}
\end{figure}

Next, we consider the entanglement generated solely due to the driven coherence by assuming atoms are injected into the cavity in their upper level.
Figure \ref{fig8} shows that the entanglement of the movable mirrors due to the driven coherence and when all atoms are injected in their upper level $|a\rangle~ (\eta=-1)$ or without injected coherence ($\rho_{ac}^{(0)}=0$). There exists a minimum strength of the cavity laser drives for which the mirror-mirror entanglement appears. The movable mirrors remain entangled for a wide range of the strength of the laser drives, with the maximum entanglement appearing at around $\Omega\approx 4.5\gamma$. The degree of the entanglement increases with increasing power of the cavity drive lasers and saturates (not shown) at $P\approx 80~\text{mW}$.

\begin{figure}
\includegraphics[width=4.2cm]{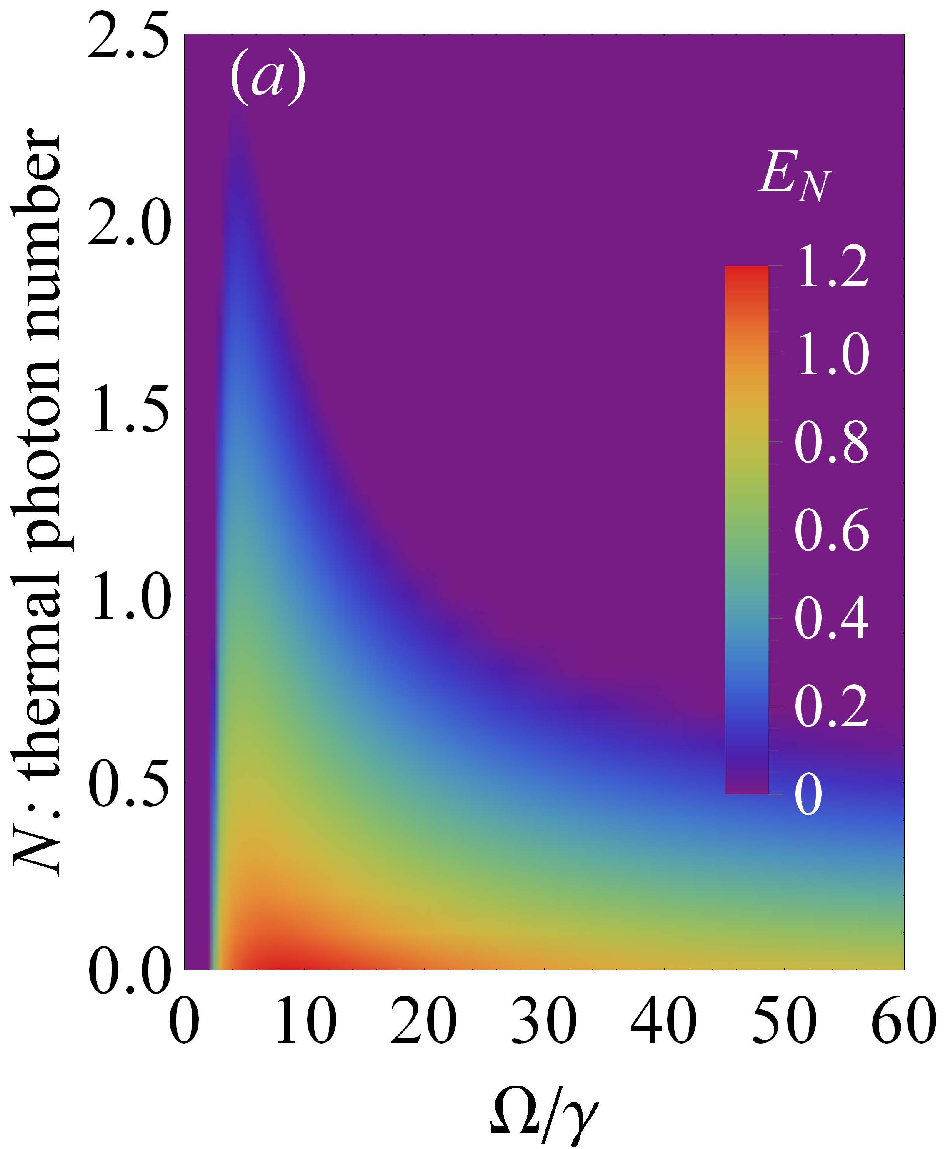}\includegraphics[width=4.05cm]{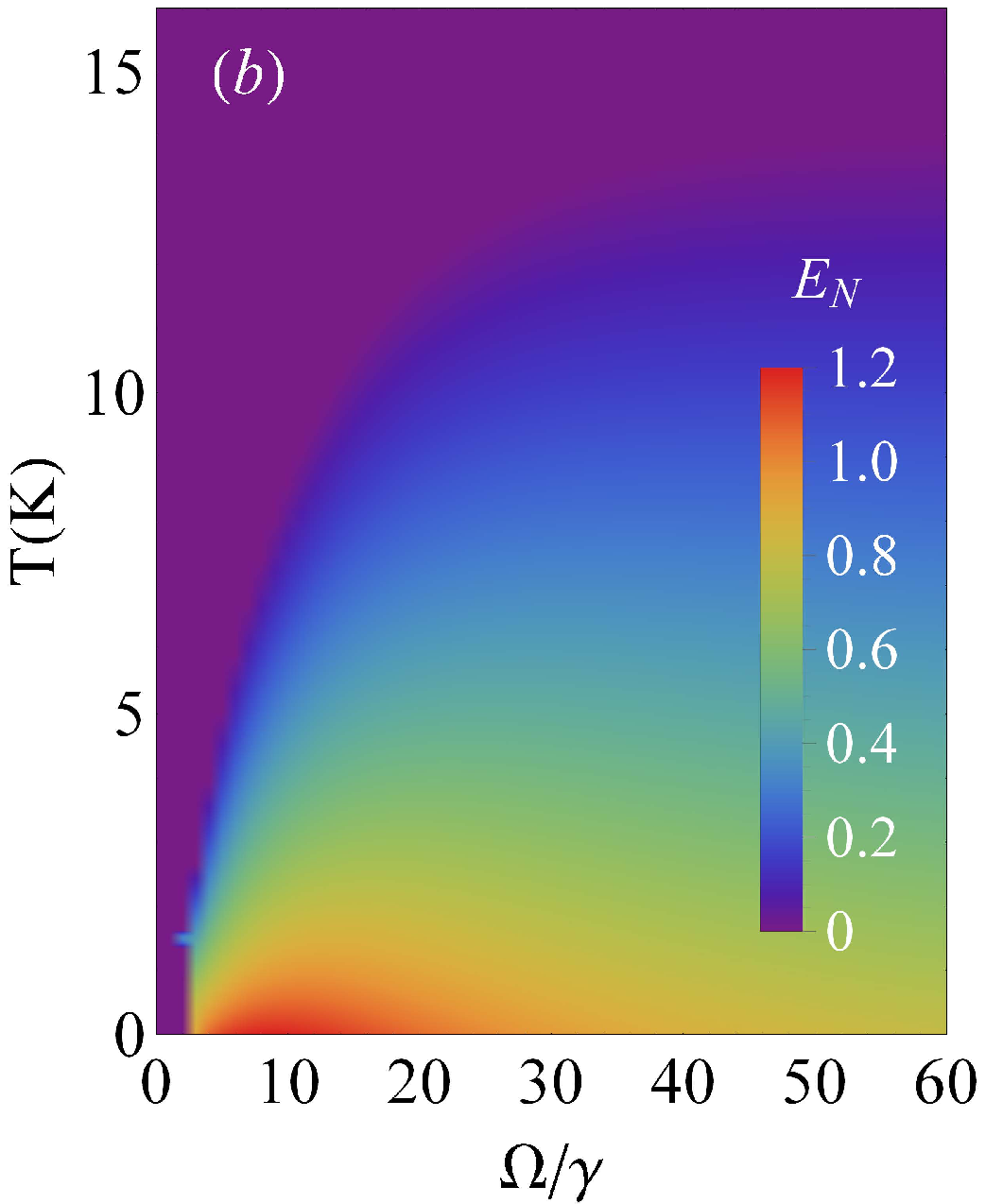}
\caption{Environment temperature dependence of the mirror-mirror entanglement with driven coherence only. (a) Logarithmic negativity $E_{N}$ vs the normalized drive amplitude $\Omega/\gamma$ of the coherent drive (for atoms) and the thermal photon numbers $N$ when the temperature of the thermal phonon bath is zero $T=0~\text{K}~(n=n_{1}=n_{2}=0)$. (b) Logarithmic negativity $E_{N}$ vs $\Omega/\gamma$ and the temperature $T$ of the thermal phonon bath when the photon bath is at zero temperature ($N=N_{1}=N_{2}=0$). The cavity drive lasers power is fixed at $P=P_{1}=P_{2}=200~\text{mW}$ and atoms injected in their upper state ($\eta=-1$). See text and Fig. \ref{fig5} for the other parameters.}\label{fig9}
\end{figure}

Finally, we studied the environmental temperature dependence of the mirror-mirror entanglement due to driven coherence and when all atoms are injected in the upper level. Our numerical simulations (see Fig. \ref{fig9}) show that at zero thermal phonon temperature and fixed cavity drive power $P=200~\text{mW}$, the entanglement decreases gradually with the number of thermal photons and eventually disappears. We note that the entanglement is more susceptible to thermal photons at higher values of the drive laser amplitude, $\Omega$. However, when the number of thermal photons is zero ($N=N_{1}=N_{2}=0$), the entanglement persists for temperature of the phonon thermal bath up to 12 K. This reveals that the entanglement generated using either injected or driven coherence disappears at the same range of phonon bath temperature.

\section{conclusion}
We analyzed the optical bistability and entanglement between two mechanical oscillators coupled to the cavity modes of a two-mode laser via radiation pressure using parameters from recent experiments. In stark contrast to the usual S-shaped bistability observed in single-mode optomechanics, we find that the optical intensities of the two cavity modes exhibit bistabilities for \textit{all values of detuning}, owing to the parametric-amplification-type coupling induced by the two-photon coherence. In addition to this, the optical intensities reveal unconventional ``ribbon''-shaped hysteresis for the circulation of optical intensities in the blue-detuned frequencies. We showed that the two-photon coherence, induced either by a strong external laser or initial preparation of the atoms of the laser medium, plays a crucial role in creating anomalous bistabilities. From application viewpoint, optical bistability has wide range potential applications from optical communications to quantum computation.

We also studied the entanglement of the movable mirrors by exploiting the intermode correlation induced by the two-photon coherence. We showed that strong mirror-mirror entanglement can be created in the adiabatic regime. Strong entanglement between the movable mirrors is obtained when the drive lasers have approximately the same power. We examined the entanglement generation due to the injected coherence and driven coherence separately. Although the two mirrors are entangled when the injected coherence is maximum, the maximum entanglement is actually achieved for slightly less coherence and when more atoms are injected in the lower level than the upper level. When the coherence is induced by a strong laser (driven coherence), there exists a threshold value of the drive strength for which the two mirrors become entangled. This entanglement then holds for wide range of the drive strength. Moreover, the entanglement created due to both coherences is remarkably robust to the phonon bath temperature, persisting up to $12~\text{K}$ for certain parameter ranges.

\begin{acknowledgements}
EAS thanks Justin Dressel and Alexander Korotkov for helpful discussions and acknowledges financial support from the Office of the Director of National Intelligence (ODNI), Intelligence
Advanced Research Projects Activity (IARPA), through the Army
Research Office Grant No. W911NF-10-1-0334. All statements of fact,
opinion or conclusions contained herein are those of the authors and
should not be construed as representing the official views or
policies of IARPA, the ODNI, or the U.S. Government. He also
acknowledge support from the ARO MURI Grant No. W911NF-11-1-0268.
\end{acknowledgements}


\begin{thebibliography}{99}

\bibitem{Man02} S. Mancini, V. Giovannetti, D. Vitali, and P. Tombesi, ``Entangling macroscopic oscillators exploiting radiation pressure,'' Phys. Rev. Lett. \textbf{88}, 120401 (2002).

\bibitem{Mar09} A. Mari and J. Eisert, ``Gently modulating optomechanical systems,'' Phys. Rev. Lett. \textbf{103}, 213603 (2009).

\bibitem{Abd12} M. Abdi, S. Pirandola,  P. Tombesi, and D. Vitali, ``Entanglement swapping with local certification: Application to remote micromechanical resonators,'' Phys. Rev. Lett. \textbf{109}, 143601 (2012).

\bibitem{Set14} E. A. Sete and H. Eleuch, ``Light-to-matter entanglement transfer in optomechanics,'' J. Opt. Soc. Am. B \textbf{31}, 2821 (2014).

\bibitem{Gua14} G. Wang, L. Huang, Y.-C. Lai, and C. Grebogi, ``Nonlinear dynamics and quantum entanglement in optomechanical systems,'' Phys. Rev.  Lett. \textbf{112}, 110406 (2014).

\bibitem{Leh13} T. A. Palomaki, J. D. Teufel, R. W. Simmonds, K. W. Lehnert, ``Entangling mechanical motion with microwave fields,'' Science \textbf{342}, 710 (2013).

\bibitem{Zha03} J. Zhang, K. Peng, and S. L. Braunstein, ``Quantum-state transfer from light to macroscopic oscillators,'' Phys. Rev. A \textbf{68}, 013808 (2003).

\bibitem{Pin05} M. Pinard, A. Dantan, D. Vitali, O. Arcizet, T. Briant, and A. Heidmann, ``Entangling movable mirrors in a double-cavity system,'' Europhys. Lett. \textbf{72}, 747 (2005).

\bibitem{Hua09} S. Huang and G. S. Agarwal, ``Entangling nanomechanical oscillators in a ring cavity by feeding squeezed light,'' New J. Phys. \textbf{11}, 103044 (2009).

\bibitem{Fab94} C. Fabre, M. Pinard, S. Bourzeix, A. Heidmann, E. Giacobino, and S. Reynaud, ``Quantum-noise reduction using a cavity with a movable mirror,'' Phys. Rev. A \textbf{49}, 1337 (1994).

\bibitem{Woo08} M. J. Woolley, A. C. Doherty, G. J. Milburn, and K. C. Schwab, ``Nanomechanical squeezing with detection via a microwave cavity,'' Phys. Rev. A \textbf{78}, 062303 (2008).

\bibitem{Set12} E. A. Sete and H. Eleuch, ``Controllable nonlinear effects in an optomechanical resonator containing a quantum well,'' Phys. Rev. A \textbf{85}, 043824 (2012).

\bibitem{Pur13} T. P. Purdy, P.-L. Yu, R.W. Peterson, N. S. Kampel, and C. A. Regal, ``Strong optomechanical squeezing of light,'' Phys. Rev. X \textbf{3}, 031012 (2013)

\bibitem{Tre96}A. Tredicucci, Y. Chen, V. Pellegrini, M. Borger, and F. Bassani, ``Optical bistability of semiconductor microcavities in the strong-coupling regime,'' Phys. Rev. \textbf{54}, 3493 (1996).

\bibitem{Dor83} A. Dorsel, J. D. McCullen, P.Meystre, E.Vignes, and H.Walther, ``Optical bistability and mirror confinement induced by Radiation pressure,'' Phys. Rev. Lett. \textbf{51}, 1550 (1983).

\bibitem{Mey85} P. Meystre, E. M. Wright, J. D. McCullen, and E. Vignes, ``Theory of radiation-pressure-driven interferometers,'' J. Opt. Soc. Am. B \textbf{2}, 1830 (1985).

\bibitem{Goz85} A. Gozzini, F. Maccarrone, F. Mango, I. Longo, and S. Barbarino, ''Light-pressure bistability at microwave frequencies,'' J. Opt. Soc. Am. B \textbf{2}, 1841 (1985).

\bibitem {Jia12} C. Jiang, B. Chen, and K.-D. Zhu, ``Controllable nonlinear responses in a cavity electromechanical system,'' J. Opt. Soc. Am. B \textbf{29}, 220 (2012).
\bibitem{Kyr13} O. Kyriienko, T. C. H. Liew, and I. A. Shelykh, ``Optomechanics with cavity polaritons: dissipative coupling and unconventional bistability,'' Phys. Rev. Lett. \textbf{112}, 076402 (2014).

\bibitem{Wei10} S. Weis, R. Rivi\'{e}re, S. Del\'{e}glise, E. Gavartin, O. Arcizet, A. Schliesser, and T. J. Kippenberg, ``Optomechanically induced transparency,'' Science\textbf{ 330}, 1520 (2010).

\bibitem{Saf11} A. H. Safavi-Naeini, T. P. Mayer Alegre, J. Chan, M. Eichenfield, M. Winger, Q. Lin, J. T. Hill, D. E. Chang, and O. Painter, ``Electromagnetically induced transparency and slow
light with optomechanics,''  Nature (London) \textbf{472}, 69 (2011).
\bibitem{Rab11} P. Rabl, ``Photon blockade effect in optomechanical systems,'' Phys. Rev. Lett. \textbf{107}, 063601 (2011).

 \bibitem{Nun11} A. Nunnenkamp, K. B{\o}rkje, and S. M. Girvin, ``Single-photon optomechanics,'' Phys. Rev. Lett. \textbf{107}, 063602 (2011).
\bibitem{Scu85} M. O. Scully, ``Correlated spontaneous-emission lasers: quenching of quantum fluctuations in the relative phase angle,'' Phys. Rev. Lett. \textbf{55}, 2802 (1985).

\bibitem{Scu82} M. O. Scully, K. Wodkiewicz, M. S. Zubairy, J. Bergou, N. Lu, and J. Meyer ter Vehn, ``Two-photon correlated-spontaneous-emission laser: quantum noise quenching and squeezing,'' Phys. Rev Lett. \textbf{60} 1832 (1988).

\bibitem{Set07} E. Alebachew, ``A degenerate three-level laser with a parametric amplifier,'' Opt. Commun. \textbf{265}, 314(2006).

\bibitem{Ale07d} E. Alebachew, ``Degenerate three-level cascade laser with the cavity mode
driven by coherent light,'' Opt. Commun.  \textbf{273}, 488 (2007).

\bibitem{Zub05} H. Xiong, M. O. Scully, and M. S. Zubairy, ``Correlated spontaneous emission laser as an entanglement amplifier,'' Phys. Rev. Lett. \textbf{94}, 023601 (2005).

\bibitem{Ale07a} E. Alebachew, ``Enhanced squeezing and entanglement in a non-degenerate three-level cascade laser with injected squeezed light,'' Opt. Commun. \textbf{280}, 133 (2007).

\bibitem{Ale07c} E. Alebachew, ``Continuous-variable entanglement in a nondegenerate three-level laser with a parametric oscillator,'' Phys. Rev. A \textbf{76}, 023808(2007).

\bibitem{Set08} E. A. Sete, ``Bright entangled light from two-mode cascade laser,'' Opt. Commun. \textbf{281}, 6124(2008).


\bibitem{Zho11} L. Zhou, Y. Han, J. Jing, and W. Zhang, ``Entanglement of nanomechanical oscillators and two-mode fields induced by atomic coherence,'' Phys. Rev. A \textbf{83}, 052117 (2011).

\bibitem{Zub13-1}W. Ge, M. Al-Amri, H. Nha, and M. S. Zubairy, ``Entanglement of movable mirrors in a correlated-emission laser,'' Phys. Rev. A \textbf{88}, 022338 (2013).

\bibitem{Zub13} W. Ge, M. Al-Amri, H. Nha, and M. S. Zubairy, ``Entanglement of movable mirrors in a correlated emission laser via cascade-driven coherence,'' Phys. Rev. A \textbf{88}, 052301 (2013).

\bibitem{Asp13} M. Aspelmeyer, T. J. Kippenberg, F. Marquardt, ``Cavity optomechanics,'' Rev. Mod. Phys. 86, 1391 (2014).

\bibitem{Scu-book97} M.O. Scully and M.S. Zubairy, \textit{Quantum Optics}, Camrigde University Press, 1997.

\bibitem{Set11b} E. A. Sete, ``Effect of dephasing on transient and steady-state entanglement in a quantum-beat laser,'' Phys. Rev. A \textbf{84},063808 (2011).

\bibitem{Wal08} D. F. Walls and G. J. Milburn, \textit{Quantum Optics}, Springer-Verlag, Berlin, 2008.

\bibitem{Ber88}J. Bergou, M. Orszag, and M. O. Scully, ``Correlated-emission laser: phase noise quenching via coherent pumping and the effect of atomic motion,'' Phys. Rev. A \textbf{38}, 768 (1988).

\bibitem{Gro09} S. Gr\"{o}blacher, K. Hammerer, M. R.Vanner, and M. Aspelmeyer, ``Observation of strong coupling between a micromechanical resonator and an optical cavity field,'' Nature (London) \textbf{460}, 724 (2009).

\bibitem{Arc06} O. Arcizet, P. -F. Cohadon, T. Briant, M. Pinard, and A. Heidmann, ``Radiation-pressure cooling and optomechanical instability of a micromirror,'' Nature (London) \textbf{444}, 71 (2006).

\bibitem{Vid02} G. Vidal and R. F. Werner, ``Computable measure of entanglement,'' Phys. Rev. A \textbf{65}, 032314 (2002).


\end{thebibliography}
 \end{document}